\documentclass[aps,showpacs,twocolumn,floatfix,amsmath,amssymb,nofootinbib]{revtex4-1}
\usepackage[export]{adjustbox}
\usepackage{amssymb}
\usepackage{amsmath}
\usepackage{epsfig}
\usepackage{graphicx}
\usepackage{latexsym}
\usepackage{bm,latexsym}
\usepackage{mathrsfs}
\usepackage{float}
\usepackage{pbox}
\usepackage[dvipsnames]{xcolor}

\def\Journal#1#2#3#4{{#1} {\bf #2}, #3 (#4)}

\def\AA{{Astron. Astrophys.}}

\def\APJL{{Astrophys. J. Lett.}}

\def\CQG{{Class. Quantum Grav.}}
\def\EPJC{{Eur. Phys. J. C}}
\def\GRG{{Gen. Relativ. Gravit.}}
\def\IJMPD{{Int. Jour. Mod. Phys. D}}
\def\JCAP{{JCAP}}

\def\LRR{{Living Rev. Rel.}} 
 
\def\NAT{{Nature}}
\def\NatAs{{Nature Astronomy}}

\def\PJP{Pramana J. Phys.}
\def\PLB{{Phys. Lett.}  B}

\def\PRL{Phys. Rev. Lett.}

\def\PRD{Phys. Rev. D}
\def\PRX{Phys. Rev. X}

\def\RMP{{Rev. Mod. Phys.}}

\begin{document}

\title{Solar system tests and chameleon effect in $f(R)$ gravity}

\author{Carolina Negrelli$^{1,2}$}
\email{cnegrelli@fcaglp.unlp.edu.ar}

\author{Lucila Kraiselburd$^{1,2}$}
\email{lkrai@fcaglp.unlp.edu.ar}

\author{Susana J. Landau$^{2,3}$}
\email{slandau@df.uba.ar}

\author{Marcelo Salgado$^{4}$}
\email{marcelo@nucleares.unam.mx}

\affiliation{$^1$Grupo de Astrof\'{i}sica, Relatividad y Cosmolog\'{i}a, Facultad de Ciencias Astron\'omicas y Geof\'{\i}sicas, 
Universidad Nacional de La Plata, Paseo del Bosque S/N (1900) La Plata, Argentina \\ $^2$CONICET, Godoy Cruz 2290, 1425 Ciudad Aut\'onoma de Buenos Aires, Argentina \\
$^3$Departamento de F\'{\i}sica, Facultad de Ciencias Exactas y Naturales, Universidad de Buenos Aires and IFIBA, 
Ciudad Universitaria - Pab. I, Buenos Aires 1428, Argentina \\ $^4$Instituto de Ciencias Nucleares, Universidad Nacional Aut\'{o}noma de M\'{e}xico,
 A.P. 70-543, M\'{e}xico D.F. 04510, M\'{e}xico.}


\date{\today}

    
\begin{abstract}
{Using a novel and self-consistent approach that avoids the scalar-tensor identification in the Einstein frame, we reanalyze the viability of $f(R)$ 
gravity within the context of solar-system tests. In order to do so, we depart from a simple but fully relativistic system of differential 
equations that describe a compact object in a static and spherically symmetric spacetime, and then make suitable linearizations that apply to non-relativistic objects such as the Sun. We then show clearly under which conditions the emerging 
{\it chameleon-like} mechanism can lead to a Post-Newtonian Parameter $\gamma$ compatible with the observational bounds. To illustrate this method, we use several 
specific $f(R)$ models proposed to explain the current acceleration of the Universe, and we show which of them are able to satisfy those bounds.}
\end{abstract}



\maketitle

\section{Introduction}
\label{sec:introduction}
$f(R)$ gravity remains one of the most popular and viable mechanisms to explain the current accelerated expansion of the Universe 
(see Refs~\cite{Sotiriou2010,Capozziello2008a,deFelice2010,Jaime2012cos,Jaime2012a,Jaime2018} for a review) while 
predicting an equation of state for the ``dark energy'' that changes in cosmic time and that might accommodate to future observations better 
than a simple cosmological constant $\Lambda$~\cite{Zhao2017,Arnold2019,DESI}. 
This proposal consists of taking for the action functional a ({\it non linear}) function $f(R)$ of the Ricci scalar $R$ 
different from the General Relativity (GR) $f_{GR}(R)= R-2\Lambda$. Thus, unless otherwise stated, and 
in order to avoid confusion, hereafter $f(R)$ refers to those non-linear models. 
This alternative, while very attractive for it does not 
require additional fields, opens, however, a Pandora box 
 that risks spoiling many of the GR predictions that have been verified with high accuracy during the past hundred years  
(e.g. solar-system tests, binary pulsar phenomenology), including the recent detection of gravitational waves 
by the LIGO-VIRGO collaboration~\cite{LIGO-VIRGO}. Several specific $f(R)$ models have been put forward to explain the cosmic acceleration, 
but many of them have failed other tests, including more refined cosmological scrutinies (e.g. the analysis of cosmological perturbations and the 
CMB), solar system (weak gravity tests) and strong gravity tests (e.g. neutron stars). One of the drawbacks of this 
kind of modifications of gravity is that there is {\it a priori} no fundamental principle that single out a $f(R)$ model\footnote{To be fair, it is important to mention that several modified theories of gravity that have been 
analyzed thoroughly in different scenarios introduce not only one arbitrary 
function but several of them. Notable examples are the generalized Galileon or Hordenski theory 
(see~\cite{Kobayashi2019} for a review) or their high-order variants~\cite{Langlois2019}, and Einstein-dilaton-Gauss-Bonnet gravity~\cite{Saffer2019} among several 
others.} (simplicity favors $f_{GR}(R)$). Moreover, among the most 
successful $f(R)$ models proposed to explain the cosmological observations there is not a single one of those that has been shown to be 
compatible with {\it all} the remaining GR tests. In particular, since the discovery of $f(R)$ theory as a potential tool 
to explain the late cosmic acceleration and the supernova Ia (SNIa) data, a controversy emerged regarding the failing of those models 
to explain the solar-system tests. The simple and heuristic argument which led to such a fallacious conclusion 
was based on the fact that one can recast $f(R)$ gravity as a kind of Brans-Dicke (BD) theory but without a kinetic term for the 
scalar field. This amounts to a BD theory with a parameter $\omega_{\rm BD}\equiv 0$. Nevertheless, we know from observations that 
$4\times 10^4 \lesssim \omega_{\rm BD}$~\cite{Alsing12}. This bound results from the relation between $\omega_{\rm BD}$ and the Post-Newtonian Parameter (PNP)
$\gamma$ which is found to be \cite{Berlotti2003}: 
\begin{equation}
\label{gamabnd}
|\gamma-1|\lesssim 2.3\times 10^{-5} \;.
\end{equation}
Therefore, the wrong and naive conclusion was that all $f(R)$ models other than $f_{GR}(R)$ are ruled out by four orders of magnitude. 
Indeed for $\omega_{\rm BD}\equiv 0$ one has $\gamma=1/2$. This confusion was clarified later by recognizing that the above conclusion 
would be valid only if the scalar-field potential that results from the identification of $f(R)$ models with 
the BD theory vanishes identically, which of course is not the case in general. Some years later, and with the advent of the so-called 
{\it chameleon theories}~\cite{Khoury2004,Burrage2018}, it was apparent that theories propagating a scalar degree of freedom (DOF) and that couples non-minimally with the matter or the curvature 
(whether the scalar-field is described in the Einstein or the Jordan frame, respectively) can produce {\it thin-shell 
effects} (screening) capable of suppressing considerably its own propagation. The point is that in {\it chameleon theories} the effective mass of the scalar-field depends on the 
density of the medium where it propagates, and so, the scalar DOF propagates differently in different media. 
Consequently,  for an $f(R)$ theory to be consistent with both cosmological and local experiments, the equivalent scalar-tensor theory must behave like a chameleon field theory.

 The screening effects depend crucially on the shape of the scalar-field potential and the detailed 
values of its parameters, which for the $f(R)$ theory at hand, this dependence translates into a specific form of this function.
For instance, it was confirmed that one of the early proposals for the late cosmic acceleration given by 
$f(R)= R - \mu^4/R$ \cite{Carroll04} is indeed unable to produce the 
screening effects, besides suffering from other problems, thus leading to $|\gamma-1|\sim 1/2$, a value which, as mentioned before, is ruled out 
by about four orders of magnitude (see Ref. \cite{Faulkner2007} for a detailed discussion on this model).

One of the difficulties in the analysis of the {\it chameleon} or {\it screening mechanism} is that it is a non-linear effect due to the 
presence of a non quadratic effective potential $V_{\rm eff}(\varphi)$ in the equation of motion for the {\it chameleon} field $\varphi$~\cite{Khoury2004,Burrage2018}. In addition, the study of such 
effect in the context of $f(R)$ gravity  complicates matters even more because the function $f(R)$ has a natural built-in scale associated with the 
cosmological distances (or equivalently, tiny cosmological densities) which contrast drastically with the solar-system scales (or densities)
which are much smaller (resp. larger) than the former. Thus a reliable analysis to solve the non-linear equation 
associated with the chameleon field requires one to handle a high numerical precision, which is usually beyond the capabilities 
of the standard number-crunching codes. Letting these technicalities aside for the moment, we feel that the analysis 
presented so far in the literature is not sufficiently convincing for a deep and simple understanding on when and how the
{\it screening mechanism} takes place in $f(R)$ gravity (cf. Section~\ref{sec:review}). Our main objection to those analyses is that they are rooted on the widespread standard method consisting of the idea that one should  {\it always} recast $f(R)$ gravity into a kind of {\it chameleon theory} in order to study the presence or not of a
{\it thin-shell} effects. Such methodology entails the following steps: 1) defining a scalar field $\chi= f_R(R)$ and inversion 
of all the variables depending on $R$ so that they depend only on $\chi$; 2) define a new {\it chameleon} like scalar-field $\varphi=\varphi(\chi)$ and a conformal 
(Einstein-frame) metric related to the original metric so that the original $f(R)$ theory looks as much as possible as the putative 
{\it chameleon theory} (with a universal coupling $\beta$). Most of the analyses presented so far use {\it mutatis mutandis} these two 
steps~\cite{Faulkner2007,Starobinsky2007,Brax2008,Capozziello08,Guo2014,Capozziello19,Katsu18,Vikram18,Hui09,Cabre12} 
among several other steps. In order to compare with the observational data one then requires to return to the original 
(Jordan frame) variables where physics finds a better and simple interpretation. 
On the other hand, very few analyses remain in the first step~\cite{Guo2014}, which amounts to dealing with a kind of hybrid (Jordan frame) scalar-tensor 
theory. 

While we do not reject the conclusions of those studies, we do believe that all such back-and-forth transformations not only 
obscure and obstruct the understanding of the predictions of solar-system observables, but are also unnecessary to find or not a
{\it screening mechanism}. 
Furthermore, many of the viable $f(R)$ models used in cosmology are not even (globally) 
invertible in terms of the variables $\varphi$ or $\chi$ and lead to potentials $V(\varphi)$ that are not single valued (see some examples in \cite{Jaime2016}).
Thus, a rigorous analysis along those lines should at least provide explicitly the domains in which such transformations are (piecewise) 
invertible. This is important because those domains may depend on the physical scenarios where $f(R)$ theory is analyzed (e.g. weak or 
strong gravity).

In order to avoid all such complications and present a straightforward and {\it cleaner} analysis, we follow the same approach that  has been 
put forward in the past to treat $f(R)$ theories \cite{Jaime2011,Jaime2013,Jaime2014,Jaime2016,Jaime2018}. This strategy consists of treating the theory directly from the 
original action functional and without performing any field redefinitions involving inversions and/or conformal transformations. 
In this way, we avoid the potential 
drawbacks described before, and in addition, we have the possibility of recovering the GR expectations in a simple manner as we will show.

Thus, our specific strategy to reanalyze the solar-system tests and the emergence of a {\it screening mechanism} 
is as follows: 1) assume $f(R)$ theory in the original variables; 2) assume a static and spherically symmetric spacetime (SSS); 
3) derive the corresponding relevant equations (for the metric and for the Ricci scalar $R$);  4) assume a matter model for the Sun 
and its outskirts, notably, the solar corona (e.g. assume a perfect fluid 
approximation and an equation of state); 5) assume that the spacetime around the solar system is a linear perturbation 
of the Minkoswki spacetime and write the equations for the metric perturbations $\phi(r)$ and $\psi(r)$; 6) perform 
suitable linear perturbations for the variable $R(r)$ around the minima of an effective potential and find the corresponding equation for the 
perturbation; 7) assume a specific $f(R)$ model and solve the resulting linear differential equations obtained in the previous steps 
under suitable boundary (regularity and asymptotic) conditions; 8) compute and compare the PNP $\gamma$ with the observational bounds.

This strategy is based on the non-perturbative (strong gravity) approach developed in Refs.
  ~\cite{Jaime2011,Jaime2018} to analyze compact objects. When adapting the matter sector to the Sun and its neighborhood,
  that approach will allow us to study systematically which $f(R)$ models are 
able to satisfy the bounds on the PNP $\gamma$ around the solar system. The paper is organized as follows. In Section~\ref{sec:f(R)} we briefly review the 
field equations of $f(R)$ theory in the (original) Jordan frame and discuss some basic properties. We also introduce some of the most popular $f(R)$ models used in the cosmological setting that we are going to analyze. 
In Section~\ref{sec:SSS} we assume a SSS spacetime to describe the spacetime around the Sun, and for the benefit of the reader,
we provide the corresponding  non-perturbative equations obtained previously in~\cite{Jaime2011,Jaime2018} for a generic $f(R)$ theory.
In Section \ref{IV}, which is the most important and novel part of the paper, we use the non-perturbative equations and perform 
a perturbative approach for the metric and the Ricci scalar indicating the precise place where the {\it chameleon like effects} 
appear and are important for recovering the observational bounds for $\gamma$.
We use these analytical tools in Section~\ref{sec:results f(R)} and confront the $f(R)$ models presented in Sec. \ref{sec:f(R)} with the 
bounds on $\gamma$. We then show which of those models are able to evade the stringent constraints placed by solar-system experiments.
Before presenting our final remarks and conclusions in
Section~\ref{sec:conclusions}, we contrast in Section~\ref{sec:review} our results with those obtained previously
in the literature in order to have an overall picture of the differences and similarities between them.


\section{$f(R)$ gravity}
\label{sec:f(R)}
The $f(R)$ theory is described by the following action functional:
\begin{equation}
\label{f(R)}
S[g_{ab},{\mbox{\boldmath{$\psi$}}}] =
\!\! \int \!\! \frac{f(R)}{2\kappa} \sqrt{-g} \: d^4 x 
+ S_{\rm matt}[g_{ab}, {\mbox{\boldmath{$\psi$}}}] \; ,
\end{equation}
where  $\kappa \equiv 8\pi G_0$ ($c=1$), $f(R)$ is an {\it a priori} arbitrary function of the Ricci scalar $R$, and ${\mbox{\boldmath{$\psi$}}}$ 
represents schematically the matter fields.

The field equation arising from variation of the action~(\ref{f(R)}) with respect to the metric is
\begin{equation}
\label{fieldeq1}
f_R R_{ab} -\frac{1}{2}fg_{ab} - 
\left(\nabla_a \nabla_b - g_{ab}\Box\right)f_R= \kappa T_{ab}\,\,,
\end{equation}
where $f_R=\partial_R f$, $\Box= g^{ab}\nabla_a\nabla_b$ is the covariant d'Alambertian and $T_{ab}$ is the energy-momentum 
tensor of matter. From this equation it is not difficult to show that $T_{ab}$ is conserved, i.e., $\nabla^a T_{ab}=0$~\cite{Koivisto2006,Jaime2016}. 

The trace of Eq.~(\ref{fieldeq1}) yields
\begin{equation}
\label{traceR}
\Box R= \frac{1}{3 f_{RR}}\left[\rule{0mm}{0.4cm}\kappa T - 3 f_{RRR} (\nabla R)^2 + 2f- Rf_R \right]\,\,\,,
\end{equation}
where $T:= T^a_{\,\,a}$. Using~(\ref{traceR}) in~(\ref{fieldeq1}) we find~\cite{Jaime2011}
\begin{eqnarray}
\label{fieldeq3}
G_{ab} &=& \frac{1}{f_R}\Bigl{[} f_{RR} \nabla_a \nabla_b R +
 f_{RRR} (\nabla_aR)(\nabla_b R) \nonumber \\
 &       &- \frac{g_{ab}}{6}\Big{(} Rf_R+ f + 2\kappa T \Big{)} 
+ \kappa T_{ab} \Bigl{]} \; ,
\end{eqnarray}
where $G_{ab}= R_{ab}-g_{ab}R/2$ is the Einstein tensor and $(\nabla R)^2:= g^{ab}(\nabla_aR)(\nabla_b R)$. 
We use Eqs.~(\ref{traceR}) and ~(\ref{fieldeq3}) as the fundamental field equations in this paper, much along the lines described in~\cite{Jaime2011,Jaime2012cos}.
 
As stressed in the Introduction, a remarkable feature of $f(R)$ theory is that it can produce in a natural fashion an accelerated 
expansion of the Universe by generating an effective cosmological constant without introducing it explicitly. We see that Eq.~(\ref{traceR}) admits $R=R_0= const$ as a particular solution when the energy-momentum tensor of matter is 
traceless ($T\equiv 0$) provided $R_0$ is an algebraic root of the implicitly defined ``potential'' ${\cal V}(R)$ via its derivative:
\begin{equation}
\frac{d{\cal V}}{dR}:= \frac{2f- Rf_R}{3f_{RR}} \; .
\end{equation}

Aside from some ``exceptional'' cases where both the numerator $2f- Rf_R$ and the denominator $f_{RR}$ vanish at $R_0$ (for example the $R^n$ model~\cite{Jaime2013}), 
in general, if $f_{RR}(R_0)\neq 0$, $R_0$ is only a root of the alternative ``potential'' (here the factor $1/3$ is kept for convention):
\begin{equation}
\label{dV}
\frac{dV}{dR}:= \frac{2f- Rf_R}{3} \; .
\end{equation}
The ``potential'' $V(R)= -R f(R)/3 + \int^R f(x) dx$ is useful to track the critical points at $R_0$, notably, the extrema (maxima or minima), at the places where its derivative (\ref{dV}) vanishes. The explicit expression for $V(R)$ (or ${\cal V}(R)$) can be computed 
once an $f(R)$ model is provided (cf. Section~\ref{sec:f(R)a}), however, even so, $V(R)$ it is not important, nor very enlightening either, as it is rather 
the derivative (\ref{dV}) which allows to locate the critical points denoted generically by $R_0$ (the reader interested in the explicit expressions 
for $V(R)$ as well as its corresponding plots for some of the models of Section~\ref{sec:f(R)a} can consult references \cite{Jaime2012cos,Jaime2012a}).
So, the three possibilities are for $R_0$ to be positive, negative or zero, which are associated with a de Sitter, anti de Sitter or Ricci flat (cosmological) background, respectively, and which give rise to an effective cosmological constant $\Lambda_{\rm eff}=R_0/4$. 
In particular, in a vacuum, the solution $R=R_0$ makes the $f(R)$ field equations to reduce to Einstein's field equations endowed with 
the above effective cosmological constant $G_{ab}= g_{ab}\frac{R_0}{4}$~\cite{fRconst,Jaime2016}. Even with the presence of matter, 
$f(R)$ models produce naturally and at late 
times the attractor solution $R\rightarrow R_0$, since, as the Universe evolves, matter dilutes with the scale factor as $1/a^3$ or $1/a^4$, i.e. 
$T_{ab}\rightarrow 0$ as $a\rightarrow \infty$, leading to an accelerated expansion 
of the Universe due to the emergence of $\Lambda_{\rm eff}$ (see, however, Ref.~\cite{Jaime2018} for an alternative possibility with a vanishing $\Lambda_{\rm eff}$).
The exact location of the critical point $R_0$ depends on the form of the $f(R)$ model and also on the specific value of the parameters involved in this function.

\subsection{$f(R)$ models}
\label{sec:f(R)a}

The $f(R)$ models considered in our analysis have to  satisfy two basic requirements: i) have theoretical consistency, such as for example the stability at the classical and semiclassical levels, and ii) be able to pass the cosmological and solar-system tests. It has been shown before that the models described below fulfill the first requirement \cite{Koyama16}. On the other hand, Jaime et al. \cite{Jaime2012cos,Jaime2012a} have  analyzed and confirmed the cosmological viability of the following models at the 
background level: the Hu-Sawicki model \cite{Hu2007}, the Starobinsky model \cite{Starobinsky2007},
and the exponential model \cite{Exponential}. They also included the logarithmic model by Miranda {\it et al.} \cite{Miranda2009} (hereafter MJWQ model), a promising model at the cosmological background level,  but which apparently suffers from several problems when analyzing the cosmological perturbations~\cite{Cruz,Miranda2}. The reason for taking into account this model is because we want to test until what extent the logarithmic models can be ruled out using the solar-system tests as claimed in~\cite{Thon}. For several $f(R)$ models, including those presented below, the predictions for the solar-system tests have also been analyzed in the scalar-tensor approach of the theory \cite{Hu2007,Thon,Cruz,Miranda2}. On the other hand, these models are built in a way that, in the high curvature regime where
$|R|\gg\breve{R}$, the resulting expression is  
$f(R)\approx R-2\Lambda_{\rm eff}^{\infty}$, where $\Lambda_{\rm eff}^{\infty}$ plays the role of an effective cosmological constant 
in that regime (e.g. $\Lambda_{\rm eff}^{\infty}=\Upsilon \breve{R}/2$, where $\breve{R}=R_s,m^2,R_{*},R_{\rm m}$ 
is a constant of the order $H_0^2$ characteristic of each $f(R)$ model described below, being $H_0$ the Hubble expansion today, 
and $\Upsilon=\lambda,1,\beta,1$ is a dimensionless constant of each model) as opposed to the effective cosmological constant
defined before $\Lambda_{\rm eff}= R_0/4$ which emerges in the low curvature regime (i.e. $R=R_0\sim H_0^2$) 
and is responsible for the late accelerated expansion of the Universe~\cite{Jaime2012cos,Jaime2012a}.

\begin{enumerate}
 
\item \textbf{The Starobinsky $f(R)$ model}
This $f(R)$ function has been proposed by Starobinsky \cite{Starobinsky2007}:

{\begin{equation}
  \label{Staromod}
f(R)=R+\lambda R_s \left[ \left(1+\frac{R^2}{R_s^2} \right)^{-q}-1\right]\;,
\end{equation}

where $R_s$, $q>0$ and $\lambda>0$ are free parameters. This model not only satisfies the necessary conditions for the existence of a viable matter-dominated epoch prior to a late-time acceleration \cite{amendola}, but also those conditions imposed by many cosmological observations such as CMB, SnIa, BAOs, cosmic chronometers, etc. (\cite{Tsujikawa08,Jaime18,PN18,Nunes,Sultana} and many others). Following \cite{Jaime2012cos} we choose  $R_s=4.17H_0^2$ and $\lambda=1$ 
in order for the model to fit the cosmological observations. 

\item \textbf{The Hu-Sawicki $f(R)$ model}

This model is defined by the function \cite{Hu2007}:
\begin{equation}
f(R)=R-m^2\frac{c_1 (R/m^2)^n}{c_2(R/m^2)^n+1}\;,
\end{equation}
where $m$, $c_1$, $c_2$ and $n>0$ are its parameters. Following \cite{Hu2007} and \cite{Jaime2012cos}  we assume $m^2=0.24H_0^2$, $c_1=1.25\times 10^{-3}$, $c_2=6.56 \times 10^{-5}$. The constants $c_1$ and $c_2$ are dimensionless and can be fixed by demanding that this model mimics as close as possible the $\Lambda\rm{CDM}$ scenario while $m^2$ has the characteristic scale of the Universe $H_0^2$. This model together with Starobinsky have been the most tested ones (\cite{Tsujikawa08,Jaime18,PN18,Nunes,Sultana,Delacruz} and many others) .

\item \textbf{Exponential $f(R)$ model}

The specific exponential model we analyze here is given by \cite{Exponential}:

\begin{equation}
f(R)=R-\beta R_{*} (1-{e^{-R/R_{*}}})\;.
\end{equation}
As in the the previous models, the parameters $R_{*}$ and $\beta$ are fixed to match the cosmological observations assuming $R_{*}=2.5H_0^2$ and $\beta=2$~\cite{Jaime2012a}. The exponential $f(R)$ has been analyzed by several authors \cite{Jaime18,Nunes,Sultana} among many others.

\item \textbf{MJWQ $f(R)$ model}

We also include the logarithmic model by Miranda {\it et al.}~\cite{Miranda2009}:

\begin{equation}
\label{fMJWQ}
f(R)=R-\alpha R_{\rm m} \ln{\left(1+\frac{R}{R_{\rm m}}\right)} \;.
\end{equation}
We have already commented that this model faces important issues with the analysis of the cosmological perturbations ~\cite{Cruz,Miranda2}. However, we include it in our analysis to  further test the limitations of the logarithmic models at 
solar-system scales~\cite{Thon}}. We take $R_{\rm m}=H_0^2$ and $\alpha=2$ which leads to a reasonable background cosmology~\cite{Miranda2009,Jaime2012cos}.
\end{enumerate}


\section{Static and spherically symmetric (SSS) spacetimes}
\label{sec:SSS}

We shall now focus on a SSS spacetime as this has proved to be a very good approximation for analyzing the solar-system tests. 
Thus we assume the following metric:
\begin{equation}
ds^2 = - n(r) dt^2  + m(r) dr^2+ r^2 d\Omega^2 \;.
\end{equation}
Using Eqs.~(\ref{traceR}) and ~(\ref{fieldeq3}) we find the required equations for the metric components 
$n(r)$ and $m(r)$, and also for the Ricci scalar~\cite{Jaime2011,Jaime2018}:

\begin{eqnarray}
\label{mprime}
m' &=& \frac{m}{r(2f_{R}+rR'f_{RR})} \Biggl{\{} 2f_{R}(1-m)-2mr^2 \kappa T^{t}_{\,\,t} 
 \nonumber \\
&&\!\!\!\!\!\!\!\!\!\!\!\!\!\!\!\!\! +\frac{mr^2}{3}(Rf_{R}+f+2\kappa T) 
+ \frac{rR'f_{RR}}{f_{R}}\Bigl{[}\frac{mr^2}{3}(2Rf_{R}-f+\kappa T) \nonumber \\
&&\!\!\!\!\!\!\!\!\!\!\!\!\!\!\!\!\!
-\kappa mr^2(T^{t}_{\,\,t}+T^{r}_{\,\,r})+2(1-m)f_{R}+2rR'f_{RR}\Bigr{]} \Biggr{\}} \; , 
\end{eqnarray}
\begin{eqnarray}
\label{nprime}
 n' &=& \frac{n}{r(2f_{R}+rR'f_{RR})} \Bigl{[} mr^2(f-Rf_{R}+2\kappa T^{r}_{\,\,r}) \nonumber \\
&& +2f_{R}(m-1)-4rR'f_{RR}  \Bigr{]} \; , \\
\label{traceRr}
R'' &=& \frac{1}{3f_{RR}}\Big{[}m(\kappa T+2f- Rf_{R}) - 3f_{RRR}R'^2\Big{]} \nonumber \\
&& +\left(\frac{m'}{2m}-\frac{n'}{2n}-\frac{2}{r}\right)R' \;.
\end{eqnarray}
These equations reduce to the corresponding GR equations for a SSS spacetime 
when $f(R)=f_{\rm GR}(R)= R-2\Lambda$.

As concerns the matter sector, we consider a perfect fluid
\begin{equation}
\label{ecu:fluido-perfecto}
T_{ab}= (\rho + p)u_a u_b + g_{ab} p \; ,
\end{equation}
where the pressure $p(r)$ and the density $\rho(r)$, are functions of the coordinate $r$ solely.

The hydrostatic equilibrium of this fluid is described by a modified Tolman-Oppenheimer-Volkoff (TOV) equation 
which arises from the conservation equation $\nabla^a T_{ab}= 0$. This equation takes the same form as in GR: 
\begin{equation}
\label{TOV}
p'= -(\rho + p) n'/2n \; .
\end{equation}
Taking into account Eq.~(\ref{nprime}), the difference of Eq.~(\ref{TOV}) with that 
of GR, is that $n'$ has additional contributions coming from the
nonlinear $f(R)$ models that have been proposed as dynamical dark-energy. Otherwise, Eq.~(\ref{TOV}) has 
exactly the same form as the TOV equation in GR.

Equation (\ref{TOV}), which describes the hydrostatic equilibrium of an object, which we will take it as the Sun, 
completes the set of differential equations. As concerns the equation of state (EOS) associated with the Hydrogen-Helium-photon content in 
the Sun we can assume different approximations. Clearly the simplest one consists of taking an {\it incompressible} fluid (i.e. constant 
density fluid) where the energy-density is given by a step function. So, the energy density $\rho$ is a nonzero constant 
$\rho_{\odot}$ within the Sun (the Sun's average density) and $\rho=\rho_{\rm cor}$ in the solar corona region. 
Outside the solar corona we assume the average density of the interstellar medium (IM) $\rho_{\rm IM}$. Thus, the total density is given by 
a {\it three step} function, each step representing the Sun's interior, the corona and the IM, respectively, with a jump discontinuity 
at the junction of the different media.

In this way, Eq.~(\ref{TOV}) can be integrated without giving any further EOS. This approximation for the EOS is sufficient to deal 
with the {\it chameleon like effects}. More detailed studies take into account more sophisticated EOS for the Sun and the corona 
\cite{Starobinsky2007,Hu2007,Guo2014}, nevertheless, as we show here, 
those details are irrelevant for recovering the solar-system tests.

The numerical integration of the equations presented in this section could be performed following the approach described in~\cite{Jaime2011}, 
which was used later in~\cite{Jaime2018}. However, 
we shall not pursue that strategy here since the numerical accuracy required to deal simultaneously with both the actual densities within the Sun, and the 
cosmological densities involved in the viable $f(R)$ cosmological models is very high and beyond the capabilities of a 
number-crunching method. Nonetheless, dealing with the full non-linear system of equations presented above represents the cleanest, 
most accurate and straightforward approach in the analysis of the solar-system tests, even if a priori we know that the Sun 
is not in the strong gravity regime. By doing so, the {\it screening} (or its absence thereof) should appear naturally in the solution depending on
the specific $f(R)$ model adopted, and then one can assess if the parameter $\gamma= |1- m(r)|/|1-n(r)|$, in the neighborhood of the 
Sun, is compatible with the observational bound (\ref{gamabnd}).

In order to avoid all the numerical complications involved in the full-fledged and non-linear treatment described in the preceding paragraph, 
and within the aim to understand in more heuristic fashion the way the chameleon effects appear, 
we shall pursue a simplified approach and perform suitable linearizations and approximations in Eqs.(\ref{mprime})$-$(\ref{traceRr}). 
Following this strategy we can find simple analytic expressions for the perturbations associated with metric components $n(r)$ and $m(r)$. 
The metric perturbations will lead to a parameter $\gamma$ which depends explicitly on 
the Ricci scalar $R$. Ultimately the analysis of $R$, which represents the scalar DOF associated with the $f(R)$ theory, will lead to a successful 
or a failure value for $\gamma$.


\section{Non-standard linear analysis}
\label{IV}
For the analysis of the solar-system tests, and more specifically, when confronting the theoretical expectations 
of the parameter $\gamma$ with the observational bounds, we will be dealing with a {\it non-standard linearization method}. This method is based on the fact that 
the metric around the Sun, as one realizes from the GR analysis, is very close to the Minkowski metric
given that the spacetime around the Sun corresponds to a weak gravitational field. Thus we define
    
\begin{eqnarray}
\label{phi}
n(r) &=& 1 - 2\phi(r) \;, \\
\label{psi}
m(r) &=& 1 + 2\psi (r) \;,
\end{eqnarray}
and assume $|\phi(r)|\ll 1$, $|\psi(r)|\ll 1$, and $|\kappa T_{ab}/\ell^2|\ll 1$, 
where $\ell \sim 150 \,\,{\rm A.U.}$ (A.U. stands for the astronomical unit) 
is a scale of the order of the solar-system size. That is, we assume that $\phi$ and $\psi$ are perturbations of the 
underlying Minkowski metric, which is assumed to be the background metric in the neighborhood of the Sun. 
Strictly speaking, the Sun is immersed in the interstellar medium (IM) which in turn is immersed within a cosmological 
background (e.g. the dark energy). All these density layers around the Sun, if considered as constant in average, 
contribute to the metric in the form of an effective ``cosmological'' constant in each substratum. In GR one usually 
ignores those contributions to the metric around the Sun given that for the IM 
$\kappa \rho_{\rm IM} r^2\sim 10^{-21} - 10^{-30}$ which is very small 
compared with $\phi(r)\sim \psi(r) \lesssim \frac{G_0 M_\odot}{{\cal R}_\odot}\times \frac{{\cal R}_\odot}{r}\sim 10^{-6}\times 10^{-4}= 10^{-10}$
in regions outside the Sun, i.e., ${\cal R}_\odot <r\lesssim 150 \, {\rm A.U.}$, where $M_\odot$ and ${\cal R}_\odot$ are the Sun's mass and radius, respectively.
The contribution of an effective dark energy is  $\Lambda_{\rm eff}\sim 10^{-30} {\rm A.U.}^{-2}$, therefore $\Lambda_{\rm eff} r^2\ll |\phi|, |\psi|$. Given these figures, current solar-system experiments are, 
in principle, unable to detect the effects of the contributions in the metric due to the IM and the dark energy as effective cosmological constants.

However, it turns out that in modified theories of gravity, especially, the ones that require a {\it chameleon-like} effect 
to suppress the scalar-degree of freedom that can potentially spoil the solar-system tests, the contributions of 
such densities are crucial. Actually, one of the key aspects of chameleon models is that the effective mass of the chameleon field depends on the density of the environments through which it propagates. Thus, a priori, one cannot neglect those densities, notably, 
the effects of the corona and the IM density in the chameleon equation, unless the model itself reveals that one can do so.

 These considerations 
lead precisely to what we consider as a non-standard linearization method. We proceed as follows. For the 
metric perturbations we keep the above prescription but in the matter terms we include the IM contributions. It is at the moment of imposing boundary conditions, namely, asymptotic conditions, that we shall 
deal with the specific asymptotic form of $\phi(r)$ and $\psi(r)$ outside the Sun. The key issue arises in the way
one treats the Ricci scalar $R(r)$ perturbatively. In the naive approach, which we include in Appendix~\ref{sec:wronglin} 
for pedagogical purposes, one assumes that $R(r)= {\tilde R}(r)+ R_0$ is a perturbed solution around just one minimum $R_0$
which corresponds to the minimum that produces an effective cosmological constant 
$R_0=\Lambda_{\rm eff}/4$. Then one linearizes Eq.~(\ref{traceRr}) for ${\tilde R} (r)$. Proceeding this way is equivalent of inhibiting all the chameleon 
effects, that is, the {\it screening} mechanism within the Sun and the corona, regardless of the specific form of 
the non-linear $f(R)$ model. The perturbation ${\tilde R} (r)$ then backreacts 
considerably in the metric perturbations, notably in $\psi(r)$, providing additional contributions which are of the same order of 
$\psi(r)$ itself leading then to an unsuitable $\gamma\approx 1/2$. 
This naive, although, inappropriate linearization scheme is the one that was considered in the early stages of the analysis of
$f(R)$ theories within the solar-system tests and which led to the (wrong) conclusion that {\it all} 
non-linear $f(R)$ models are ruled out (cf. Appendix~\ref{sec:wronglin}).

Owing to this drawback, we are forced to follow a different strategy. The idea is that when taking into account the Sun, the corona and the IM,
the three media will produce a different effective scalar-field potential for $R$, each one with its corresponding minimum at $R_{\rm min}^{\rm in,cor,IM}$. 
Thus, in order to take into account the {\it chameleon-like} effect, and if a linear method proves to be valid, 
 we require to linearize Eq.~(\ref{traceRr}) around each minimum, i.e., around three 
non-perturbed backgrounds (the Sun's interior, the corona, and the IM), which is a quite non-standard method. 
Notwithstanding, it is important to stress that this method is not totally novel since 
in the original chameleon model a similar {\it linearization} method that takes 
into account different media has proved to be valid in certain regimes (e.g. the thin and thick shell approximations)
\cite{Khoury2004,Kraiselburd2018}. The difference here is that we are taking into account the backreaction of the {\it chameleon like effects} on the gravitational field generated by the Sun. Actually, the perturbation ${\tilde R} (r)$ that allows one to interpolate $R(r)$ 
between the three  minima, $R_{\rm min}^{\rm in,cor,IM}$, will not be necessarily a {\it small} perturbation relative to $R_{\rm min}^{\rm in,cor,IM}$, as the three minima can be very different
from each other precisely due to the contribution of the three completely different values of the densities to the 
three effective scalar-field potentials. 
This is precisely one of the complications that emerges when trying to analyze chameleon models, which are 
inherently non-linear ones, with linear approximations. We can, however, proceed with 
this non-standard linearization method as long as we implement it with care. In fact, if the chameleon effect ensues, the interpolation between 
the three minima occurs basically within confined thin-shells of very small size. Thus, even if the error on 
the approximate solution for $R(r)$ committed in these shells turns out to be large, the error remains in these narrow 
shells. Different linear approximations for the effective potentials can be implemented in order to 
decrease that error. In particular, due to the screening effects, one expects that inside the Sun and the corona
  the Ricci scalar will remain very close to their minima, i.e., $R_I(r)\approx R_{\rm min}^I$
  (where $I={\rm in,cor}$) except perhaps within a (narrow) region near the edge of the Sun and the corona where
  $R (r)$ interpolates between the two minima, $R_{\rm min}^{\rm in}$ and  $R_{\rm min}^{\rm cor}$. On the other hand, in the IM we expect $R(r)\approx R_{\rm min}^{\rm IM} + {\tilde R}_{\rm IM}(r)$, where 
$|{\tilde R}_{\rm IM}(r)/R_{\rm min}^{\rm IM}|\ll |\psi|,|\phi|$ if the screening happens. 
It is the quantity ${\tilde R}_{\rm IM}(r)$ that can contribute to the metric potentials outside the Sun, but if suppressed, it will not spoil 
the observational bounds for  $\gamma$.
On the other hand,   {\it chameleon-like} screening mechanisms depend  on the medium density. Therefore, an appropriate description of the solar corona should be included when considering solar-system tests. Thus, we  stress that we take into account the corona not because we want to model the bending of light due to the optical effects produced by the media (i.e. refraction), but because the corona can increase the {\it chameleon-like} effect and mitigate the abrupt decrease of densities (between the Sun's interior and the IM).
The corona helps to smooth further the transition of the field $R(r)$ between the Sun's interior and the IM. We checked that if the corona is not included in the analysis the screening is less effective 
and may lead to a value of $\gamma$ that is ruled out by observations. 
Moreover, the effects near the Sun's surface may be important since the observational value of $\gamma$ is measured when the Earth and Saturn are in conjunction.

Consequently, we consider perturbations inside the Sun, inside the Sun's corona and outside the corona around the location of the respective minima 
$R_{\rm min}^{\rm in, cor, IM}$ associated with the respective effective potentials (see Section~\ref{sec:QA} below):
\begin{equation}
\label{Rtilde}
R(r)=
\begin{cases}
 R_{\rm in}(r)= R_{\rm min}^{\rm in} + {\tilde R}_{\rm in} (r)\hspace{0.6cm} (0\leq r \leq {\cal R}_{\odot}) \\
\\
 R_{\rm cor}(r)= R_{\rm min}^{\rm cor} + {\tilde R}_{\rm cor} (r)\hspace{0.6cm} ({\cal R}_{\odot} \leq r \leq {\cal R}_{\rm cor}) 
\\
\\
R_{\rm IM}(r)= R_{\rm min}^{\rm IM} + {\tilde R}_{\rm IM}(r) \hspace{0.5cm}  ({\cal R}_{\rm cor} \leq r \leq {\cal R}_{\rm IM})
\end{cases}
\end{equation}
where we take ${\cal R}_{\rm cor} = 15$ ${\cal R}_\odot$, ${\cal R}_{\rm IM}\sim 150$ ${\rm A.U.}$, and assume  ${\tilde R}_{\rm in} (r) \ll R_{\rm min}^{\rm in}$, ${\tilde R}_{\rm IM} (r) \ll R_{\rm min}^{\rm IM}$, and ${\tilde R}_{\rm cor} (r) \ll R_{\rm min}^{\rm cor }$. 

In Section~\ref{sec:QA} we solve Eq.(\ref{traceRr}) perturbatively for each ${\tilde R}_{\rm in,cor,IM}(r)$ 
(the Sun's interior, the corona and the IM region) and then match continuously the three solutions at the transition zones 
$r= {\cal R}_{\odot}$ and $r={\cal R}_{\rm cor}$ in order to obtain a full solution $R(r)$, within the three layers. 
Our goal is then to analyze the extent to which the  backreaction of  $R(r)$ into the metric potentials $\phi(r)$ and $\psi(r)$
 (at this level of approximations) and for some specific non-linear $f(R)$ models, 
leads, due to the {\it screening,} to the value $\gamma\sim 1$ required by the observations in the solar system. 
As shown in Section \ref{sec:QA} the {\it screening} can occur but, unlike the original chameleon model~\cite{Khoury2004}, it is basically due to an exponential 
suppression outside the Sun rather than due to a {\it thin-shell parameter}. Thus, in this paper, 
and under the way we treat $f(R)$ theories, we allude to a {\it screening} rather than to a thin-shell effect  \footnote{In the original chameleon model~\cite{Khoury2004}, the scalar-field $\varphi(r)$ outside a high density spherical body of radius ${\cal R}$ 
has the following form $\varphi_{\rm out}(r)=\varphi_{\rm min}^{\rm out}+ {\rm const}\times \frac{\Delta{\cal R}}{{\cal R}}\times \frac{e^{-mr}}{r}$ (being $m=m_{\varphi,{\rm eff}}^{\rm out}$). In that model when the screening effects take place, 
the following conditions occur, $m_{\varphi,{\rm eff}}^{\rm out} r \ll 1$ and $\frac{\Delta{\cal R}}{{\cal R}}\ll 1$.
    Thus the screening is mainly due to the {\it thin-shell} parameter $\frac{\Delta{\cal R}}{{\cal R}}$ and thus $\varphi_{\rm out}(r)\approx \varphi_{\rm min}^{\rm out}$. 
Under the present approach, outside the Sun, notably, in the
    corona region, ${\tilde R}_{\rm cor}(r)$ behaves like  $|\varphi_{\rm out}(r)-\varphi_{\rm min}^{\rm out}|$ except that the coefficient that multiplies the Yukawa term is not necessarily ``small'', however,
    the quantity $m_{\rm eff}^{\rm cor} r $ is sufficiently large to suppress ${\tilde R}_{\rm cor}(r)$ at the place it reaches the IM layer. 
See Section~\ref{sec:QA} for the details.}.

As stressed above, in order to deal with Eq.~(\ref{traceRr}) we will adopt 
additional approximations. For instance, we will neglect the contributions of the curved spacetime only in this equation. 
That is, in that equation we take $n(r)= 1$, and $m(r)= 1$. The reason behind this assumption is because we know that the {\it screening, chameleon-like effect} emerges from the behavior of ${R} (r)$ itself and not from the metric perturbations. This is a fact 
that is observed in the original chameleon analysis proposed by Khoury \& Weltman\cite{Khoury2004} as well as in many other subsequent 
investigations~(see~\cite{Burrage2018} for a review), 
where one neglects the backreaction of the metric into the chameleon field. This is presumably a good approximation in the {\it weak field} regime. 
Moreover, this approximation is also consistent with the linearization method that we outlined previously, and therefore 
Eq.~(\ref{traceRr}), which is associated with the perturbation ${\tilde R}(r)$ (in its three layers), is treated in a flat spacetime background.
Indeed, the most important features to take into account in this 
equation are the effective potentials inside the Sun, and outside (the corona and the IM), which depend on the details involved in the function 
$f(R)$ itself and on the densities of the Sun, and the two outer layers. \footnote{We take $n(r)= 1$, and $m(r)= 1$ {\it only} in the equation for  $R(r)$. 
Clearly, one requires $n= 1-2\phi$ and 
$m=1+2\psi$ in order to produce a non-zero $R(r)$, if the Ricci scalar is computed directly from the metric. At this point we emphasize that 
if Eqs. ~(\ref{mprime}), ~(\ref{nprime}) for $m'$ and $n'$, respectively, together with an equation for $n''$ (not shown here; see~\cite{Jaime2011}) 
are used to compute $R$ from its explicit expression from the metric one finds $R\equiv R$, showing the self-consistency of the method~\cite{Jaime2011}. 
The linear approximation for $R$ in terms of $\phi$ and $\psi$ will contain terms linear in $\psi''$ and $\phi''$.
 Thus, in $f(R)$ gravity one can obtain $R$ from the metric or from (\ref{traceRr}) which was already used in obtaining the first order equations 
~(\ref{mprime}) and~(\ref{nprime}). Like in previous studies~\cite{Jaime2011,Jaime2016,Jaime2018} we use (\ref{traceRr}) to compute $R$.}

Thus, we first proceed to insert Eqs.~(\ref{phi})--(\ref{Rtilde}) into Eqs.~(\ref{mprime})--(\ref{traceRr}) and keep only the terms linear in 
$\phi$, $\psi$, $\tilde R$ and linear in the matter terms $T_{ab}$. 
We remind the reader that by following this approximation we are assuming that the metric around the Sun is very close to the 
Minkowski metric and that $\phi$, $\psi$ are only metric perturbations, much as one does in GR when taking the weak field limit. Thus, we assume 
$|{\tilde R}_I/R_{\rm min}^I|\sim |\phi|\sim |\psi|\sim \kappa \rho r^2$ where the script $I$ stands for ``in'', ``cor'' and ``IM''
(see Section~\ref{sec:QA} for the details on the total $\rho$). If our previous assumptions were invalid, that is, 
$|\psi|\ll |{\tilde R}_I/R_{\rm min}^I|$ or $|\phi|\ll |{\tilde R}_I/R_{\rm min}^I|$,  
one could not expect a weak field limit in the solar system, which does not seem a very appealing situation. 

\subsection{Linearization of metric perturbations}

A straightforward calculation leads to the following linearized equation for $\psi$: 

\begin{eqnarray}
\label{psiprime}
\psi' &\approx & \frac{1}{4 r f_{R_{\rm min}} }\Big\{ -4\psi f_{R_{\rm min}} - 2\kappa r^2 T^{t}_{\,\,t} \nonumber \\
&& + \frac{r^2}{3} \Big[ (1+2\psi)(R_{\rm min} f_{R_{\rm min}} + f_{\rm min})  \nonumber \\
&&  + {\tilde R}\Big(R_{\rm min} f_{RR_{\rm min}} + 2f_{R_{\rm min}}\Big)
+ 2\kappa T \Big] \nonumber \\
&& 
 +  \frac{r^3 {\tilde R}^\prime}{3}\frac{f_{RR_{\rm min}}}{f_{R_{\rm min}}}\left(2 R_{\rm min} f_{R_{\rm min}}-f_{\rm min}\right)
\Big\} \;,
\end{eqnarray}

where $R_{\rm min}$ corresponds to the location of the minima (if they exist) of the effective potential $V_{\rm eff}(R,T)$
inside and outside the Sun satisfying 
\begin{equation}
\label{R0cond}
\frac{dV_{\rm eff}}{dR}:= \frac{\kappa T + 2f - R f_{R}}{3f_{RR}}\Big{|}_{R_{\rm min}}=0 \;.
\end{equation}
In Eqs.~(\ref{psiprime}) and (\ref{R0cond}) all the quantities are evaluated inside and outside the Sun, but for brevity we omit the ``in'', ``cor'' and ``IM'' labels, although later we will be more explicit in this matter. 
Since we will assume $f_{RR}|_{R_{\rm min}}\neq 0$ (see below), then Eq.~(\ref{R0cond}) reduces to 

\begin{equation}
\label{minconpot}
\kappa T + 2f - R f_{R}\Big{|}_{R_{\rm min}}=0 \;.
\end{equation}
At the minima, the effective mass associated with $R$ is $m_{\rm eff}^2:= d^2V_{\rm eff}/dR^2|_{R_{\rm min}} >0$. Thus,
\begin{eqnarray}
\label{meff1}
m_{\rm eff}^2 &=& \frac{f_R-Rf_{RR}}{3f_{RR}}\Big{|}_{R_{\rm min}} = \frac{\kappa T+ 2f- R^2f_{RR}}{3Rf_{RR}} \Big{|}_{R_{\rm min}}.
\end{eqnarray}

We appreciate that the effective mass $m_{\rm eff}$ depends explicitly on the 
density $\rho\approx - T$ of each of the three media
\footnote{Alternatively, 
the effective mass associated with the field $\chi= f_R$ (see Eq.~(\ref{tracefRrap}) below) proposed in other treatments~\cite{Guo2014}, also depends on the density.}. 
One requires $f_R>0$, notably at the minimum, in order to have 
  a positive effective gravitational constant $G_{\rm eff}= G_0/f_R$. Moreover, for weak fields one expects
  $R\sim \kappa \rho$, and so, $R\geq 0$. Thus, in order to have a positive $m_{\rm eff}^2$, assuming, $R_{\rm min}>0$ and $f_R>0$,
  one demands the following two conditions: $f_{RR}|_{R_{\rm min}}>0$ and $Rf_{RR}|_{R_{\rm min}}< f_R |_{R_{\rm min}}$.
Specific $f(R)$ models not satisfying these conditions seem unsuitable to produce a physically reasonable scenario, whether 
at the solar system level or as a cosmological model. For instance the model $f(R)= R - \mu^4/R$ has $f_{RR}= -2\mu^4/R^3$ which is 
negative for $R>0$, and it has an effective potential with extrema at $R_{\pm}= 
\frac{1}{2}[\kappa\rho \pm \sqrt{(\kappa\rho)^2 + 12\mu^4}]$ and $m_{\rm eff}^2= -\frac{R_{\pm}^3}{6\mu^4}(1+3\mu^4/R_{\pm}^2)$. In particular, 
in regions where the density vanishes (vacuum) $R_{\pm}= \pm \mu^2\sqrt{3}$. The value $R_{+}>0$ is required for the model to generate a
positive $\Lambda_{\rm eff}$, and thus, a late accelerated expansion of the Universe. Then taking $R_{\rm +}= \mu^2\sqrt{3}$
leads to $m_{\rm eff}^2= -3\mu^2/\sqrt{3}$.
Due to the problematic result $m_{\rm eff}^2<0$, this model is not only prone to tachyonic instabilities~\cite{Dolgov2003}, 
but also unable to exhibit the {\it screening effect} that is needed to suppress the influence of the scalar-degree of freedom on 
the metric, and thus, it is unable to pass the solar-system tests.

At this point it is worth remarking that in GR $f_{RR}\equiv 0$, $f_{R}\equiv 1$ and $R=-\kappa T$. In such a
case Eq.~(\ref{psiprime}) is replaced by\footnote{\label{foot:GRpsi}In principle one cannot take the GR limit directly from Eq.~(\ref{psiprime})
  as we considered perturbations about the minimum $R_{\rm min}$ which assumes $f_{RR}|_{R_{\rm min}} \neq 0$. However, formally
  one can take such a limit in the constant density scenario if $R=R_{\min}= -\kappa T\approx \kappa \rho=const$ and ${\tilde R}\equiv 0$, that is,
  assuming that $R$ is a step function proportional to the density, and neglecting the term $r^2 \psi R_{\min}$ relative to
  $\psi$: $r^2 \psi R_{\min}\sim -r^2 \psi
  \kappa T \sim r^2 \psi \kappa \rho_{\odot} = \psi\times \frac{3\kappa M_{\odot}}{4\pi {\cal R}_\odot}\times \frac{r^2}{{\cal R}_\odot^2}=
  \psi \times \frac{6 G_0 M_{\odot}}{\cal{R}_\odot}\times \frac{r^2}{{\cal R}_\odot^2} \ll \psi$.}
 \begin{equation}
\label{psiGR}
\psi' \approx -\frac{\psi}{r} - \frac{\kappa r T^{t}_{\,\,t} }{2}= -\frac{\psi}{r} + \frac{\kappa r \rho }{2}  \;.
\end{equation}

The vacuum (exterior) solution of this equation is simply $\psi_{\rm ext}(r)\approx const/r$, which when matching with the interior solution $\psi_{\rm in} (r)
\approx G_0 M_{\odot} r^2/{R^3_\odot}$ at $r={\cal R}_\odot$ (where $M_\odot= 4\pi \rho_\odot {\cal R}_\odot^3/3$) one obtains 
$\psi_{\rm ext} (r)\approx G_0 M_{\odot}/r$, which is the well-known weak-field-limit solution  in vacuum
around the Sun. If one includes the IM in the GR solution one has $\psi_{\rm ext} (r)\approx G_0 M_{\odot}/r + \kappa \rho_{\rm IM}r^2/6$. Assuming $\kappa \rho_{\rm IM}\sim 10^6 \Lambda_{\rm eff}$ (i.e. $\rho_{\rm IM}\sim 10^6 \rho_{\rm cosmo}$), the second term 
is still ten orders of magnitude smaller than the first one in the neighborhood of the Sun, so we can neglect it as well, as
we mentioned earlier.

The most important aspect of our analysis is when $f(R)\neq f_{GR}(R)$ and thus, we have to take into account 
the terms with ${\tilde R}^\prime$ and ${\tilde R}$ in Eq.~(\ref{psiprime}). 

An interesting aspect of Eq.~(\ref{psiprime}), as well as its non-linear version Eq.~(\ref{mprime}), is that it is completely 
decoupled from $n$ and thus, from $\phi$. Eq.~(\ref{psiprime}) is coupled to the scalar DOF of the theory represented by ${\tilde R}$, 
and is also coupled to the matter terms. However, since we shall assume a perfect and non relativistic incompressible fluid for the Sun, 
for the Sun's corona and for the IM, we shall have $T^{t}_{\,\,t}= -\rho= const$ and $T= -\rho+ 3p \approx -\rho= const$ (except for the discontinuities
of the density at $r={\cal R}_\odot$ and ${r=\cal R}_{\rm cor}$). Hence, the matter sector will be also decoupled 
from this metric perturbation and we do not need to solve for the matter part within the incompressible-fluid approximation. The price to pay is that 
$\psi$ will have a caustic at $r={\cal R}_\odot$ and ${r=\cal R}_{\rm cor}$ (i.e. $\psi$ is continuous but not differentiable at those places) where $\rho$ experience a jump. This drawback 
happens as well in GR when using an incompressible fluid for the Sun 
and will not affect our conclusions.

Let us now consider Eq.~(\ref{psiprime}) inside the Sun, and use (\ref{minconpot}) in this region. 

If within the Sun the following conditions are verified 
\begin{eqnarray}
  \label{fminaproxsun}
  f_{\rm min}\sim R_{\rm min} \;, \\
  \label{fRminaproxsun}
  f_{R_{\rm min}}\sim 1 \;,
\end{eqnarray}
then from Eq.~(\ref{minconpot}) one expects
\begin{equation}
  \label{Rmininaprox}
R_{\rm min} \sim - \kappa T_\odot\approx \kappa \rho_\odot \;.
\end{equation}
Moreover,  $-\kappa  T_\odot r^2 \sim \frac{3\kappa M_\odot}{4\pi {\cal R}_\odot}\frac{r^2}{{\cal R}_\odot^2}
=\frac{6 G_0 M_\odot}{{\cal R}_\odot} \frac{r^2}{{\cal R}_\odot^2}
\sim \psi_{\rm in}$. That is, inside the Sun $\psi\sim \kappa \rho_\odot r^2$ and from Eq.(\ref{Rmininaprox}) we conclude $\psi_{\rm in} \sim R_{\rm min}^{\rm in} r^2$. The same approximation holds outside the Sun, given that 
$-\kappa T_{\rm cor,IM} r^2\sim \kappa \rho_{\rm cor,IM}r^2 \sim \psi_{\rm cor,IM}|_{r\sim {\cal R}_{\rm cor,IM}}$. 
If conditions similar to (\ref{fminaproxsun})--(\ref{Rmininaprox}) hold outside the Sun (with $\rho_\odot$ replaced by 
$\rho_{\rm cor}$ or $\rho_{\rm IM}$), then 
$\psi_{\rm cor} \sim \frac{GM_{\odot}}{R_\odot}\times \frac{R_\odot}{r} + R_{\rm min}^{\rm cor} r^2 \lesssim 
\frac{GM_{\odot}}{R_\odot} + R_{\rm min}^{\rm cor} r^2 \sim \psi_{\rm in}({\cal R}_\odot) + R_{\rm min}^{\rm cor} r^2$ and similarly
$\psi_{\rm IM} \lesssim \psi_{\rm in}({\cal R}_\odot) + R_{\rm min}^{\rm IM} r^2$. As a consequence for  $\psi_{\rm in,cor,IM}$ the term 
$ 2\psi r^2 (R_{\rm min} f_{R_{\rm min}} + f_{\rm min})\sim  4r^2  R_{\rm min} \psi \sim 4 r^2 \kappa \rho \psi\sim 4 \times {\cal O}(\psi^2)$. 
In this way Eq.~(\ref{psiprime}) can be approximated everywhere by: 
\begin{eqnarray}
\label{psiprimesun2}
\psi' &\approx & -\frac{\psi}{r} - \frac{\kappa r T^{t}_{\,\,t}}{2 f_{R_{\rm min}} } \nonumber \\
&& + \frac{r}{12  f_{R_{\rm min}} }\Big[ (R_{\rm min} f_{R_{\rm min}} + f_{\rm min}) + 2\kappa T  \nonumber \\
&&  + {\tilde R}\Big(R_{\rm min}f_{RR_{\rm min}} + 2f_{R_{\rm min}}\Big)  \Big]\nonumber \\
&&  +  \frac{r^2 {\tilde R}^\prime}{12}\frac{f_{RR_{\rm min}} }{f_{R_{\rm min}}}\left(2R_{\rm min}-\frac{f_{\rm min}}{f_{R_{\rm min}}}\right) \;,
\end{eqnarray}
where for brevity we have not included the ``in'', ``cor'' and ``IM'' labels in the above equation.

Using Eq.~(\ref{minconpot}) in (\ref{psiprimesun2}) we obtain
\begin{eqnarray}
\label{psiprimesun3}
\psi' &\approx & -\frac{\psi}{r}+ \frac{r}{4 f_{R_{\rm min}} } \Big( - 2\kappa T^{t}_{\,\,t} + \kappa T + f_{\rm min}\Big) \nonumber \\
&& + \frac{1}{12  r f_{R_{\rm min}} } \Big[ {\tilde R}r^2 \Big( R_{\rm min}f_{RR_{\rm min}} + 2f_{R_{\rm min}}\Big)  \nonumber \\
&&  + \frac{r^3 {\tilde R}^\prime f_{RR_{\rm min}}}{f_{R_{\rm min}}} \left(2 R_{\rm min} f_{R_{\rm min}}-f_{\rm min}\right) \Big] \;,
\end{eqnarray}
which can be written as 
\begin{eqnarray}
\label{psiprimesun4}
\frac{1}{r}\frac{d (r\psi)}{dr} &\approx & \frac{r}{4 f_{R_{\rm min}} } \Big( \kappa \rho + f_{\rm min}\Big) \nonumber \\
&& + \frac{1}{12  r f_{R_{\rm min}} } \Big[ {\tilde R}r^2 \Big( R_{\rm min}f_{RR_{\rm min}} + 2f_{R_{\rm min}}\Big)  \nonumber \\
&&  + \frac{r^3 {\tilde R}^\prime f_{RR_{\rm min}}}{f_{R_{\rm min}}} \left(2 R_{\rm min} f_{R_{\rm min}}-f_{\rm min}\right) \Big] \;,
\end{eqnarray}
where as before we used  $T^{t}_{\,\,t} = -\rho$ and $T= - \rho+ 3p\approx - \rho$.

It is important to provide some insight about the final expectations concerning the chameleon or screening effects when the latter take place
within some of the specific non-linear $f(R)$ models. In order to do so, and for the sake of the following heuristic 
argument, we do not take into account the solar corona and consider only two regions: the 
interior of the Sun and the IM as a single environment. If the {\it chameleon-like} effects were {\it ideal}, then within
  the Sun $R_{\rm in}(r)\approx R_{\rm min}^{\rm in}$, and outside $R_{\rm out}(r)\approx R_{\rm min}^{\rm out}$
  (with $R_{\rm min}^{\rm in,out}\approx -\kappa T_{\rm in,out}\approx \kappa \rho_{\rm in, out}$), except within
a very narrow region at $r={\cal R}_\odot$ where $R(r)$ would experience a sharp decreasing (almost a discontinuous jump). 
Under the current approach we consider that $R(r)$ is $C^1$
at the edges, so ${\tilde R}$ is not exactly zero everywhere, although it is sharply peaked within a narrow region near the edges of the layers. 
In principle, these narrow regions can be made arbitrarily small in an ideal screening mechanism, in which case ${\tilde R}$
    becomes like a Dirac delta, and so ${\tilde R}\approx 0$ inside and outside the Sun except near the edge 
and as we approach this limit the full solution for $R(r)$ becomes an almost perfect  step function with
the ``step'' localized at $r={\cal R}_\odot$.

In that case the interior solution of Eq.~(\ref{psiprimesun4}) is given simply by
\begin{equation}
  \psi_{\rm in}(r)\approx \frac{r^2}{12 f_{R_{\rm min,in}}}\Big(\kappa \rho_\odot + f_{\rm min,in}\Big) \;\;\; (0\leq r\leq {\cal R}_\odot)
  \end{equation}
where, like in the GR case, the integration constant was set to zero imposing regularity at the origin  $r=0$, i.e. that  the metric component $m(0)=1= 1+2\psi (0)$, and so $\psi (0)=0$ [cf. Eq.~(\ref{psi})].

On the other hand, outside the Sun
\begin{eqnarray}
  \psi_{\rm out}(r) && \approx \frac{r^2}{12 f_{R_{\rm min,out}}}\Big(\kappa \rho_{\rm IM} + f_{\rm min,out}\Big)\nonumber \\
  && + \frac{const}{r} \;\;\; ({\cal R}_\odot\leq r\leq {\cal R}_{\rm IM}) \,.
  \end{eqnarray}
If one simply neglects the contribution of the IM in the exterior solution (where
$f_{\rm min, out}\approx \kappa \rho_{\rm IM}$), then
\begin{equation}
  \psi_{\rm out}(r) \approx \frac{const}{r} \;\;\; ({\cal R}_\odot\leq r\leq {\cal R}_{\rm IM}) \:.
  \end{equation}
Matching the interior and the exterior solutions at $r={\cal R}_\odot$, one finds
\begin{equation}
  const = \frac{G_0 M_\odot}{2 f_{R_{\rm min,in}}}\left( 1 + \frac{ f_{\rm min,in}}{\kappa \rho_\odot}\right).
\end{equation}
Furthermore, in an ideal screening scenario $f_{\rm min,in}= R_{\rm min,in}\approx \kappa \rho_\odot$ and $f_{R_{\rm min,in}}= 1$ then
$const \approx G_0 M_\odot$, and so 
\begin{equation}
  \psi_{\rm out}(r) \approx \frac{G_0 M_\odot}{r} \;\;\; ({\cal R}_\odot\leq r\leq {\cal R}_{\rm IM})
\end{equation}
and we recover the GR expectations for $\psi$. We shall illustrate later that under such approximations $\phi_{\rm out}\approx \psi_{\rm out}$
and the PNP $\gamma\approx 1$ is recovered. Thus, we conclude that an ideal screening allows us to recover the GR limit.

Notwithstanding, and within the context of a more realistic model for the Sun that we put forward, which includes the solar corona, 
the {\it screening effect} is not an ideal one, but it has to ensure that, while
$R(r)$ is not exactly a perfect step function, and so ${\tilde R}(r)$ is not exactly null, 
${\tilde R}(r)$ must be very small almost everywhere (i.e. ${\tilde R}(r)\ll R_{\rm min}$, except
perhaps within  narrow regions near $r={\cal R}_\odot$ and $r={\cal R}_{\rm cor}$). In such an instance,
the effects of the scalar DOF become sufficiently small so that it backreacts very weakly in the solutions for $\psi$ and $\phi$. 
The effective mass, $m_{\rm eff}^{\rm in,cor,IM}$ (\ref{meff1}) is the quantity that modulates the behavior of  ${\tilde R}(r)$ in the different layers. 

As concerns the linearization of Eq.~(\ref{nprime}), we proceed in a similar fashion as for the linearization of Eq.~(\ref{mprime})
and obtain
\begin{eqnarray}
\label{phiprime}
\phi' &\approx& -\frac{\psi}{r} - \frac{1}{4rf_{R_{\rm min}}}\Big\{r^2(1+2\psi)\Big(f_{\rm min}-R_{\rm min}f_{R_{\rm min}}\Big)\nonumber \\
&&  -r^2 {\tilde R} R_{\rm min}f_{RR_{\rm min}} -4rf_{RR_{\rm min}} {\tilde R}^\prime + 2\kappa r^2 T^{r}_{\,\,r}\Big\} \;.
\end{eqnarray}

If conditions similar to (\ref{fminaproxsun})--(\ref{Rmininaprox}) hold inside the three regions the previous equation can be approximated by:

\begin{eqnarray}
\label{phiprime2}
\phi' &\approx& -\frac{\psi}{r} - \frac{r}{4f_{R_{\rm min}}}\Big(f_{\rm min}-R_{\rm min}f_{R_{\rm min}} + 2\kappa T^{r}_{\,\,r} \Big) \nonumber \\
&& + \frac{f_{RR_{\rm min}}}{4rf_{R_{\rm min}}} \Big(r^2{\tilde R} R_{\rm min} + 4r {\tilde R}^\prime\Big) \;,
\end{eqnarray}
where we used $R_{\rm min}\sim -\kappa T$, and so $r^2 R_{\rm min}\sim {\cal O}(\psi)$,
and, thus, in the right-hand-side (r.h.s) of Eq.~(\ref{phiprime}) we neglected a term $r^2\psi (f_{\rm min}-R_{\rm min} f_{R_{\rm min}})
= r^2\psi R_{\rm min}(f_{\rm min}/R_{\rm min} - f_{R_{\rm min}})\sim \psi \times {\cal O}(\psi) (f_{\rm min}/R_{\rm min} - f_{R_{\rm min}})$ 
since it is small compared with $\psi$. The factor $(f_{\rm min}/R_{\rm min} - f_{R_{\rm min}})$ can make the previous term even smaller
if $f_{\rm min}\sim R_{\rm min}$ and $f_{R_{\rm min}}\sim 1$.

Using Eq.~(\ref{minconpot}) in Eq.~(\ref{phiprime2}) yields
\begin{eqnarray}
\label{phiprime4}
\phi' &\approx& -\frac{\psi}{r} - \frac{ r }{4f_{R_{\rm min}}}\Big(2\kappa T^{r}_{\,\,r}- \kappa T - f_{\rm min}\Big)\nonumber \\
&& + \frac{f_{RR_{\rm min}}}{ f_{R_{\rm min}}}\left({\tilde R}^\prime + \frac{r R_{\rm min} {\tilde R}}{4} \right) 
\;.
\end{eqnarray}
Taking $T^{r}_{\,\,r}= p$ we find 
\begin{eqnarray}
\label{phiprime5}
\phi' &\approx&  -\frac{\psi}{r} - \frac{ r }{4f_{R_{\rm min}}}\Big(2\kappa p + \kappa \rho - f_{\rm min}\Big)\nonumber \\
&& + \frac{f_{RR_{\rm min}}}{ f_{R_{\rm min}}}\left({\tilde R}^\prime + \frac{r R_{\rm min} {\tilde R}}{4} \right)\;.
\end{eqnarray}
In GR Eq.~(\ref{phiprime4}) or (\ref{phiprime5}) reduces to \footnote{Differentiating Eq.~(\ref{phiaproxGR}) and using (\ref{psiGR}) 
and Eq.~(\ref{phiaproxGR}) again 
we obtain  $\frac{1}{r^2}\frac{d}{dr}\left(r^2 \phi'\right)= \nabla^2 \phi = -\frac{\kappa \rho}{2}= -4\pi G_0 \rho$, where $\nabla^2$
  stands for the Laplacian in spherical coordinates. In this way we recover the Newtonian equation for the gravitational potential $\Phi=-\phi(r)$.}
\begin{eqnarray}
\label{phiaproxGR}
\phi' &\approx& -\frac{\psi}{r} - \frac{\kappa r T^{r}_{\,\,r}}{2} \approx -\frac{\psi}{r} \;,
\end{eqnarray}
when neglecting the pressure term $T^{r}_{\,\,r}=p$. 

Unlike $\psi'$, we appreciate from Eq.~(\ref{phiaproxGR}) that the derivative $\phi'$ is continuous at $r={\cal R}_\odot$ and ${r=\cal R}_{\rm cor}$
  even in the incompressible fluid approximation because in this case $\psi$ is continuous. For instance,
  at $r={\cal R}_\odot$, $\phi'|_{{\cal R}_\odot}\approx -\frac{\psi}{r} |_{{\cal R}_\odot}$, which is well defined even if the density experiences a discontinuous jump at $r={\cal R}_\odot$
(i.e. $\phi$ turns out to be $C^1$ at ${\cal R}_\odot$).  Clearly the exterior solution of Eq.~(\ref{phiaproxGR}) 
in vacuum is  $\phi_{\rm ext} (r)\approx G_0 M_{\odot}/r = \psi_{\rm ext}$. Thus we recover the PNP $\gamma= |\psi_{\rm ext}/\phi_{\rm ext}| \approx 1 + {\cal O}(\delta)$, where $\delta\lesssim 10^{-6}$  stands for $\phi$ and $\psi$ (i.e. the $\delta$ 
corrections appear when taking into account the quadratic corrections in the metric). 
Since we shall impose that ${\tilde R}$ is at least 
$C^1$ at ${\cal R}_\odot$, we will not encounter a discontinuity in $\phi$ either even when we assume the non linear models  $f(R)\neq f_{GR}(R)$
and thus, when we take into account the contributions of the scalar DOF ${\tilde R}$. 

We can subtract Eqs.~(\ref{psiprimesun3}) and ~(\ref{phiprime4}) and obtain
\begin{eqnarray}
\label{psimphi}
&& \frac{d}{dr}\left(\psi-\phi +  \frac{{\tilde R} f_{RR_{\rm min}}}{f_{R_{\rm min}}}\right) \approx
\frac{\kappa r }{2f_{R_{\rm min}}}\left(T^{r}_{\,\,r}- T^{t}_{\,\,t}\right)
\nonumber \\
&& + \frac{r {\tilde R}}{6}\left(1-\frac{R_{\rm min} f_{RR_{\rm min}}}{f_{R_{\rm min}}}\right) \nonumber\\ 
&& + \frac{r^2 {\tilde R}^\prime f_{RR_{\rm min}}}{12 f_{R_{\rm min}}^2} \left(2 R_{\rm min} f_{R_{\rm min}}-f_{\rm min}\right)\;.
\end{eqnarray}
In the GR case this equation reduces to
\begin{eqnarray}
\label{psimphiGR}
 \frac{d}{dr}\left(\psi-\phi\right) \approx \frac{\kappa r }{2}\left(T^{r}_{\,\,r}- T^{t}_{\,\,t}\right)\;.
\end{eqnarray}
In vacuum this equation leads to $\phi_{\rm ext} (r)= \psi_{\rm ext} (r)$ as before\footnote{In the full non-linear GR case one obtains 
$nm=-g_{tt}g_{rr}=1$ when 
the EMT satisfies the condition $T^{r}_{\,\,r}= T^{t}_{\,\,t}$ (see Ref.\cite{Salgado2003} for a
thorough discussion). This result can be extended to modified metric theories of gravity if the theory can be written as 
$G_{ab}= \kappa T_{ab}^{\rm eff}$ and if and only if the effective EMT of the underlying theory verifies the conditions
$T^{r}_{{\rm eff},\,\,r}= T^{t}_{{\rm eff},\,\,t}$ in area coordinates of SSS spacetimes.}. For the interior and exterior solutions $T^{r}_{\,\,r}= p$ and $T^{t}_{\,\,t}= -\rho$ 
and for a non-relativistic fluid such as the Sun, the corona and the IM, $p\ll \rho$, thus we can neglect the pressure term\footnote{The pressure term will be relevant 
only in the TOV equation because it is the pressure that maintains the hydrostatic equilibrium, within the Sun in this case. For our purposes it is not relevant to take into 
account this equation because, as stressed before, the equations for $\psi$, $\phi$ and ${\tilde R}$ are not coupled to $p$, but only to $\rho$ 
when assuming the non-relativistic condition $p\ll \rho$, as well as the constant density fluid.}.

Next we argue why part of the second term in Eq.~(\ref{psimphi}) can be neglected 
everywhere. For this, we analyze the dimensionless quantity $R_{\rm min} f_{RR_{\rm min}}/f_{R_{\rm min}}$. We are interested in testing 
$f(R)$ models that were proved to be cosmologically viable. Those models 
have a built-in scale $R_*\sim H_0^2$ and thus, have the form $f(R)= R + R_* F_1(\frac{{R}}{R_*})$, where $F_1$ is a dimensionless function of its argument.
 Thus, $f_{R_{\rm min}}= 1 + dF_1/dz|_{{z_{\rm min}}}$ where $z= R/R_*$ and
$f_{RR_{\rm min}}= \frac{1}{R_*}\frac{d^2 F_1}{dz^2}|_{{z_{\rm min}}} $   Consequently,
$\frac{R_{\rm min} f_{RR_{\rm min}}}{f_{R_{\rm min}}}=\frac{R_{\rm min}}{R_*} \frac{d^2 F_1}{dz^2}|_{z_{\rm min}} \Big(\frac{1}{1 + dF_1/dz|_{z_{\rm min}}}\Big)$ and we have checked that 
  for the  cosmologically viable $f(R)$ models considered in this paper the following two conditions are satisfied 
$\frac{dF_1}{dz}|_{{z_{\rm min}}}\ll 1$, and $\frac{R_{\rm min} f_{RR_{\rm min}}}{f_{R_{\rm min}}} \approx z_{\rm min}\times \frac{d^2 F_1}{dz^2}|_{z_{\rm min}}\ll 1$ 
(see Table~\ref{TabfRR}). 

\begin{table}[!hbtp]
\centering
\begin{tabular}{|l|l|l|l|}
\hline
\textbf{$f(R)$}   &  \textbf{Sun} &  \textbf{Corona} &  \textbf{\rm IM}    \\ \hline
Starobinsky $q=2$ & $5.6\times10^{-145}$ & $3.35 \times 10^{-69}$ & $3.35 \times 10^{-24}$ \\ \hline
Starobinsky $q=0.4$ & $2.3\times10^{-52}$ & $4.33 \times 10^{-25}$ & $7.01 \times 10^{-9}$ \\ \hline
Hu-Sawicki $n=4$ & $1.06\times10^{-145}$ & $6.36 \times 10^{-70}$ & $6.36 \times 10^{-25}$ \\ \hline
MJWQ & $3.64\times10^{-30}$ & $6.23 \times 10^{-19}$ & $5.34 \times 10^{-6}$ \\ \hline
\end{tabular}
\caption{Values of the dimensionless quantity $z \frac{d^2 F_1}{dz^2}|_{{z_{\rm min}}}$ 
evaluated in the three media of our Sun's model (columns 2--4) for the four non-linear $f(R)$ models (column 1) 
described in Sec.~\ref{sec:f(R)a}}.
\label{TabfRR}
\end{table}

Now, let us analyze the third line of Eq.~(\ref{psimphi}) in the IM, for simplicity, although numerically 
we will take into account the corona as well. We can write:
 \begin{eqnarray}
  \Sigma &:=& \frac{r^2 {\tilde R}^\prime_{\rm IM} f_{RR_{\rm min}}}{12 f_{R_{\rm min}}^2} \left(2 R_{\rm min} f_{R_{\rm min}}-f_{\rm min}\right)
\Big|_{R_{\rm min}=R_{\rm min}^{\rm IM}} \nonumber \\
   && = \frac{r^2 R_{\rm min} f_{RR_{\rm min}}}{6 f_{R_{\rm min}}}{\tilde R}^\prime_{\rm IM}\left(1-\frac{f_{\rm min}}{2R_{\rm min}f_{R_{\rm min}} }
   \right)\Big|_{R_{\rm min}=R_{\rm min}^{\rm IM}} \nonumber \\
   && =\Big(\frac{r}{{\cal R}_{\rm IM}}\Big)^2 \times {\cal R}_{\rm IM}^2 R_{\rm min}\times R_{\rm min} f_{RR_{\rm min}} \times \frac{{\tilde R}^\prime_{\rm IM}}{R_{\rm min}} \nonumber \\
 &&  \times \left(1-\frac{f_{\rm min}}{2R_{\rm min} f_{R_{\rm min}}}\right) \Big|_{R_{\rm min}=R_{\rm min}^{\rm IM}}.
 \end{eqnarray}

By the same arguments given above, $f_{\rm min}= R_{\rm min}\Big[1 + \frac{R_*}{R_{\rm min}} F_1(z_{\rm min})\Big]=R_{\rm min}\Big[1 + \frac{F_1(z_{\rm min})}{z_{\rm min}} \Big]$. We computed the quantity $F_1(z_{\rm min})/z_{\rm min}$ in the IM 
for the four $f(R)$ models considered in this paper (see Secs.~\ref{sec:f(R)a} and \ref{sec:results f(R)}) 
and found that it is always very small (see Table \ref{TabfRR2}), and 
$f_{R_{\rm min}^{\rm IM}}\approx 1$. Therefore we conclude that $f_{\rm min,IM}\approx R_{\rm min}^{\rm  IM}$ and
$\left(1-\frac{f_{\rm min}}{2R_{\rm min}f_{R_{\rm min}} }\right) \Big|_{R_{\rm min}=R_{\rm min}^{\rm IM}}\lesssim 1$. Moreover,
 ${\cal R}_{\rm IM}^2 R_{\rm min}^{\rm IM}\sim  {\cal R}_{\rm IM}^2 \kappa\rho_{\rm IM}\sim 10^{-21}$.  Finally, 
 $R_{\rm min}^{\rm IM} f_{RR_{\rm min}^{\rm IM}} \times \frac{{\tilde R}^\prime}{R_{\rm min}^{\rm IM}} \ll \frac{{\tilde R}^\prime}{R_{\rm min}^{\rm IM}}$
 since, as we showed,  $R_{\rm min}^{\rm IM} f_{RR_{\rm min}^{\rm IM}}\ll 1$. We thus conclude
 \begin{equation}
   \Sigma = \lambda \Big(\frac{r}{{\cal R}_{\rm IM}}\Big)^2 \times \frac{{\tilde R}^\prime_{\rm IM}}{R_{\rm min}^{\rm IM}} \;,
 \end{equation}
where $\lambda< 10^{-21}$ is a dimensionless coefficient.

\begin{table}[!hbtp]
\centering
\begin{tabular}{|l|l|}
\hline
\textbf{$f(R)$}   &  \textbf{\rm IM}   \\ \hline
Starobinsky $q=2$ & $1.1\times10^{-5}$  \\ \hline
Starobinsky $q=0.4$ & $4.0\times10^{-14}$\\ \hline
Hu-Sawicki $n=4$ & $1.2\times10^{-5}$ \\ \hline
MJWQ & $6.8\times10^{-5}$\\ \hline
Exponential & $1.3\times10^{-5}$\\ \hline
\end{tabular}
\caption{Dimensionless quantity $F_1(z_{\rm min})/z_{\rm min}$ evaluated in the interstellar medium 
(second column) for the non-linear $f(R)$ models described in Sec.~\ref{sec:f(R)a} (first column).}
\label{TabfRR2}
\end{table}
So in the IM Eq.~(\ref{psimphi}) can be very well approximated by
\begin{eqnarray}
\label{psimphi2}
&& \frac{d}{dr}\left(\psi_{\rm IM} -\phi_{\rm IM} +  \frac{\tilde R_{\rm IM}}{R_*}\frac{d^2F_1}{dz^2}|_{z_{\rm min}^{\rm IM}}\right)\nonumber \\
&&\approx \frac{r}{6}\left(3\kappa \rho_{\rm IM} +
      {\tilde R}_{\rm IM} \right) + \lambda \Big(\frac{r}{{\cal R}_{\rm IM}}\Big)^2 \frac{{\tilde R}^\prime_{\rm IM}}{R_{\rm min}^{\rm IM}} \nonumber \\
&&      = \frac{3\kappa \rho_{\rm IM} r}{2} \left(1 + \frac{{\tilde R}_{\rm IM}}{3 \kappa \rho_{\rm IM}} \right)
+ \lambda \Big(\frac{r}{{\cal R}_{\rm IM}}\Big)^2 \frac{{\tilde R}^\prime_{\rm IM}}{R_{\rm min}^{\rm IM}}.
\end{eqnarray}

Now, if $R_{\rm min}^{\rm IM}\approx \kappa \rho_{\rm IM}$, and by hypothesis ${\tilde R_{\rm IM}}\ll R_{\rm min}^{\rm IM}$, we obtain 
\begin{eqnarray}
\label{psimphi3}
&&
\frac{d}{dr}\left(\psi_{\rm IM} -\phi_{\rm IM} +  \frac{\tilde R_{\rm IM}}{R_*}\frac{d^2F_1}{dz^2}|_{z_{\rm min}^{\rm IM}}\right) \nonumber \\
&& \approx \frac{3\kappa \rho_{\rm IM} r}{2} + \lambda \Big(\frac{r}{{\cal R}_{\rm IM}}\Big)^2 \frac{{\tilde R}^\prime_{\rm IM}}{R_{\rm min}^{\rm IM}}\;.
\end{eqnarray}
Integrating this equation from $r={\cal R}_{\rm cor}$ to $r\leq {\cal R}_{\rm IM}\sim 150$ A.U. we find
\begin{eqnarray}
\label{psimphiint}
&& \psi_{\rm IM} (r)  \approx \phi_{\rm IM}(r) -  \frac{\tilde R_{\rm IM} (r)}{R_*}\frac{d^2F_1}{dz^2}|_{z_{\rm min}^{\rm IM}} \nonumber \\
&& +  \frac{3\kappa \rho_{\rm IM} {\cal R}_{\rm IM}^2}{4}\left(1 - \frac{r^2}{{\cal R}_{\rm IM}^2} \right)
+ \lambda \Big(\frac{r}{{\cal R}_{\rm IM}}\Big)^2 \frac{{\tilde R}_{\rm IM}(r)}{R_{\rm min}^{\rm IM}}\nonumber \\
&& - \lambda \Big(\frac{{\cal R}_{\rm cor}}{{\cal R}_{\rm IM}}\Big)^2 \frac{{\tilde R}_{\rm IM}({\cal R}_{\rm cor})}{R_{\rm min}^{\rm IM}}
 -\frac{2\lambda}{{\cal R}_{\rm IM}^2} \int_{{\cal R}_{\rm cor}}^r  \frac{x{\tilde R}_{\rm IM}(x)}{R_{\rm min}^{\rm IM}} dx \nonumber \\
&& + const \;,  \;\;\;\;\;\;\;\;\;\;\;\;\;\;\;\;\;\;\;\;\;\;\;\;\;\;\; ({\cal R}_{\rm cor}\leq r\leq {\cal R}_{\rm IM})
\end{eqnarray}
where the integration constant is fixed by demanding 
  $\psi_{\rm IM} \approx  \phi_{\rm IM}$ at $r= {\cal R}_{\rm IM}$. Previously we showed that $\kappa \rho_{\rm IM} {\cal R}_{\rm IM}^2 \sim 10^{-21}$, and since
$\frac{{\cal R}^2_{\rm cor}}{{\cal R}_{\rm IM}^2}<\frac{r^2}{{\cal R}_{\rm IM}^2} \leq 1$ in the IM,
we conclude that the term involving $\rho_{\rm IM}$ is negligible. Moreover, we also expect that $\tilde R_{\rm out}({\cal R}_{\rm IM})\approx 0$ 
so that $R({\cal R}_{\rm IM})\approx R_{\rm min}^{\rm IM}$, and by hypothesis ${\tilde R}_{\rm IM}(r)/R_{\rm min}^{\rm IM}\ll 1$.
Finally, $r/{\cal R}_{\rm IM}\leq 1$. All these considerations imply that the terms multiplying $\lambda$, including the
term involving the integral, are very small compared with $\psi_{\rm IM}(r)$ and
$\phi_{\rm IM}(r)$, both of which are of the order $G_0 M_{\odot}/{\cal R}_\odot \sim 10^{-6}$. In fact, using the mean-value theorem it is not difficult to see that
the term involving the integral is bounded by $2\lambda$ and $\lambda\lesssim 10^{-21}$. So we obtain

\begin{equation}
\label{psimphiout1}
\psi_{\rm IM} (r)  \approx  \phi_{\rm IM}(r) - \frac{\tilde R_{\rm IM} (r)}{R_*}\frac{d^2F_1}{dz^2}|_{z_{\rm min}^{\rm IM}} + const  \;.
\end{equation}
From the condition $\psi_{\rm IM}({\cal R}_{\rm IM}) \approx  \phi_{\rm IM}({\cal R}_{\rm IM})$, we conclude that the integration constant is negligible as well.
Thus, from Eq.(\ref{psimphiout1}) it is possible to find a solution for $\phi_{\rm IM}(r)$ in terms of the additional perturbations: 
\begin{eqnarray}
&& \phi_{\rm IM} (r)  \approx \psi_{\rm IM}(r) +  \frac{\tilde R_{\rm IM} (r)}{R_*}\frac{d^2F_1}{dz^2}|_{z_{\rm min}^{\rm IM}} \nonumber \\
\label{psimphiout}
&& =\psi_{\rm IM}(r) +  \frac{\tilde R_{\rm IM} (r)}{R_{\rm min}^{\rm IM}}\times z_{\rm min}^{\rm IM}\times  \frac{d^2F_1}{dz^2}|_{z_{\rm min}^{\rm IM}} \;.
\end{eqnarray}
The perturbation $\psi_{\rm IM}(r)$ will be computed from Eq.~(\ref{psiprimesun4}).
  The second term on the r.h.s. of Eq.~(\ref{psimphiout}) is precisely the one that includes
the {\it chameleon like effects} or {\it screening}, depending on the behavior of ${\tilde R}_{\rm IM}(r)$.
So, when the {\it screening} effects take place, the Ricci scalar perturbation ${\tilde R}_{\rm IM}(r)$ is suppressed by the Yukawa behavior 
associated with $m_{\rm eff}^{\rm IM}$ (see Section~\ref{sec:QA} below) and so the contribution to the metric perturbations due to ${\tilde R}_{\rm IM}(r)$
becomes very small, and the chameleon effect of the theory allows one to recover the GR expectations. Moreover, 
according to Table~\ref{TabfRR} the factor $z_{\rm min}^{\rm IM}\times \frac{d^2F_1}{dz^2}|_{z_{\rm min}^{\rm IM}}\ll 1$,  
produces an additional suppression of the scalar DOF on the metric perturbations. In reality, and for some 
$f(R)$ models, some of the terms that involve the coefficient $\lambda$ in Eq.~(\ref{psimphiint}) might be of the same order of magnitude 
as the second term on the r.h.s of Eq.~(\ref{psimphiout}). Nevertheless, all such terms turn out to be very small when the screening 
occurs. For instance, one can take additional terms proportional to ${\tilde R}_{\rm IM}(r)$ [cf. Eqs.~(\ref{phioutcor}) and (\ref{phiout})].

However, for $f(R)$ models not satisfying the previous considerations, the second term on the r.h.s of Eq.~(\ref{psimphiout}) is of order
$\psi_{\rm IM}(r)$ and $\phi_{\rm IM}(r)$, and thus, the scalar DOF destroys the GR predictions implying the unviability of those specific modified-gravity models.

In Sec~\ref{sec:QA} below, we perform an analysis
similar to Eq.~(\ref{psimphi}) taking into account the corona region ${\cal R}_\odot\leq r\leq {\cal R}_{\rm cor}$ 
as well. Moreover, in order to complete the analysis of the full system of linearized equations we need to study the linearization method for Eq.~(\ref{traceRr}).
  In the following sections we present the linear approximation of this equation, and thus, analyze the perturbation  $\tilde R (r)$,
  its solution and the way the latter backreacts on the metric. Then we provide numerical results for the quantity $\gamma(r) := \psi (r)/\phi (r)$
  outside the Sun and confront them with the observational bound (\ref{gamabnd})  for several of the cosmological viable $f(R)$ models.

\subsection{Linearization of the Ricci scalar equation}
\label{sec:QA}

First, taking Eq.~(\ref{traceRr}) in a Minkowski background yields
\begin{equation}
\label{traceRrap}
R'' +\frac{2 R'}{r} = \frac{1}{3f_{RR}}\Big{(}\kappa T+2f- Rf_{R} - 3f_{RRR}R'^2\Big{)}  \;,
\end{equation}
where we omit for the moment the labels ``in, cor, IM'' associated with the solutions in those regions.

In most treatments\cite{Faulkner2007,Guo2014}, this equation is written in terms of the scalar field $\chi=f_R$:
\begin{equation}
\label{tracefRrap}
\chi'' +\frac{2 \chi'}{r} = \frac{1}{3}\Big{(}\kappa T+2f- R\chi  \Big{)} \;,
\end{equation}
where $R$ and $f(R)$ are implicit functions of $\chi$. However, the disadvantage of using Eq.~(\ref{tracefRrap}) over Eq.~(\ref{traceRrap}) is that we have to invert all functions of $R$ in terms of $\chi$ which is
possible globally (i.e. in the entire domain for $R$) provided $f_{RR}\neq 0$, namely $f_{RR}>0$, in order to avoid instabilities~\cite{Dolgov2003}. 
This condition does not hold globally in all cosmologically viable $f(R)$ models like in the Hu-Sawicki~\cite{Hu2007} and
  Starobinsky~\cite{Starobinsky2007} models but only piece-wisely (cf. \cite{Jaime2016}). We shall then deal with Eq.~(\ref{traceRrap}). After all, 
  the {\it chameleon effect} is a physical one and cannot depend on the change of variables, particularly 
in cases where the transformation from $R$ to $\chi$ is well defined globally.

Except for the term involving ${R}'^2$, Eq.~(\ref{traceRrap}) has a structure very similar to that of the original chameleon equation. For instance, the terms 
$\kappa T/f_{RR}$ and $(2f- Rf_{R})/f_{RR}$ in Eq.~(\ref{traceRrap}) play the same role as the terms $-\sqrt{\kappa}\beta T\varphi $ and $dV/d\varphi$,
respectively, of the original chameleon equation~\cite{Khoury2004}, where $dV/d\varphi$ is defined in terms of the fundamental (bare) potential of the theory.
 Thus, the key part in the analysis for the recovering of the solar-system tests is a detailed study of Eq.~(\ref{traceRrap}) and, in particular, the way $R(r)$ backreacts on the perturbed metric. Although the approximate Eq.~(\ref{traceRrap}) still is non-linear 
and it can only be solved numerically in general, like it happens in the original chameleon model~\cite{Khoury2004},  
we can approximate it linearly by following the non-standard approach described before: we linearize Eq.~(\ref{traceRrap}) around the 
minima associated with the effective potential $V_{\rm eff}^{\rm in,cor,IM}$ in  each media, 
where $V_{\rm eff}^{\rm in,cor,IM}$ satisfies  Eq.~(\ref{R0cond}) with $f_{RR}|_{R_{\rm min}^{\rm in,cor,IM}}\neq 0$, notably $f_{RR}|_{R_{\rm min}^{\rm in,cor,IM}}>0$. That is, in each region we approximate the 
effective potential quadratically around each minima, which amounts to approximate Eq.~(\ref{traceRrap}) linearly around the 
corresponding minima. We then solve the resulting equations in each region and then match continuously the solutions at the border of each 
layer. This is a good approximation provided $R_{\rm in,cor,IM}(r)\approx R_{\rm min}^{\rm in,cor,IM}= const$, in most of the regions  
considered, notably, in the regions where the solar-system tests take place.
Since the main errors committed on the total solution are to be confined near the boundary layers, or in regions where the 
solution is already small, the global solution is reliable, particularly, for the application of those tests. 
In the past that method has been applied to the original chameleon equation~\cite{Kraiselburd2018} providing 
accurate results, notably in the thin-shell regime.

There exists also a linearization method that consists of linearizing Eq.~(\ref{traceRrap}) around a unique cosmological background
$R_0$, however, this method is flawed as it does not allow for the screening effects to appear (see Appendix~\ref{sec:wronglin}).

As mentioned in Section~\ref{IV}, the corona has to be taken into account in the theoretical model as it can affect the analysis of the chameleon-like effect. Hence, we consider a model with three layers: the Sun, its corona and the interstellar medium, each of them with constant densities,  as follows:
 \begin{equation}
\rho(r)=
\begin{cases}
  \rho_{\odot}= 1.43 \,\,{\rm g  \,\,cm}^{-3}\hspace{0.9cm} (0\leq r \leq {\cal R}_\odot) \\ \\
\rho_{\rm cor}= 10^{-15} \,\,{\rm g\,\, cm}^{-3}\hspace{0.8cm} ({\cal R}_\odot\leq r \leq {\cal R}_{\rm cor}) \\  \\
\rho_{\rm IM}= 10^{-24} \,\, {\rm g \,\, cm}^{-3} \hspace{0.8cm}  ({\cal R}_{\rm cor} \leq r \leq {\cal R}_{\rm IM})
\end{cases}
\label{rho}
\end{equation}
Our model is simpler than others proposed in the literature \cite{Starobinsky2007,Hu2007,Guo2014}, where the density is not taken 
to be homogeneous within the different layers. This simplification allows us to obtain an analytic solution for  Eq.(\ref{traceRrap}), which it is not possible 
with a varying density model. 

Nevertheless, in order to test the approximation (\ref{rho}) considered here, we also compute the value of $\gamma$ 
by varying the size of the corona\footnote{We remind the reader that in our analysis we take ${\cal R}_{\rm cor} = 15 {\cal R}_{\odot}$ which is a quite conservative value. }, ${\cal R}_{\rm cor} = 2 {\cal R}_{\odot}, 7 {\cal R}_{\odot}, 14 {\cal R}_{\odot}$, but 
taking the same density, and find no significant differences in our results. Therefore, we believe that whether a particular $f(R)$ model passes (or not) the 
solar-system tests is not due to the use of this simplified model for the Sun
  and its neighborhood, and so, the same conclusion is expected to hold if a more accurate model for the matter part is employed. In other words, from our analysis we have evidence that detailed matter models for the Sun and the corona are not crucial for the recovering or not of the screening effects.

Proceeding with the linearization of Eq.~(\ref{traceRrap}) around the three minima we have,
\begin{equation}
\label{eqRL}    
R''_{I}+\frac{2 R'_{I}}{r}\approx m_{{\rm eff}, I}^2(R_{I}-R_{\rm min}^{I}) \;,
\end{equation}
or equivalently
\begin{equation}
\label{eqRLtilde}    
{\tilde R}''_{I}+\frac{2{\tilde R}'_{I}}{r}\approx m_{{\rm eff}, I}^2{\tilde R}_{I} \;,
\end{equation}
where, as before, the index ${I}$ stands for ``in'', ``cor'' and ``IM'', and the effective mass is given by Eq.~(\ref{meff1}).

For the above equation to be valid inside the Sun, the $f(R)$ theory under consideration must meet the following condition (see the Appendix~\ref{appA} for the details): 

\begin{equation}
  \label{TSP}
\frac{(R_{\rm min}^{\rm in}-R_{\rm min}^{\rm cor})f_{RR_{TS}}}{ G_0 M_\odot  /{\cal R}_\odot}\ll 1.
\end{equation}
where $R_{TS}$ is the solution of the equality $m_{\rm eff}^2(R-R_{\rm min})=\frac{-\kappa\rho}{3f_{RR}}$ inside the Sun. Also, it is necessary that the Compton condition  $L\gg (m_{\rm eff}^{\rm cor})^{-1}$ ($L$ is the width of the corona region) is fulfilled inside the Sun's corona. If this condition  is not satisfied,  a new effective mass for the corona must be defined and the appropriate $R_{\rm min}^{\rm cor}$  which is no longer a minimum should be calculated. In the next subsection we will mention for which particular model this last case applies.

\begin{figure}[t]
\centering
\includegraphics[width=0.52\textwidth]{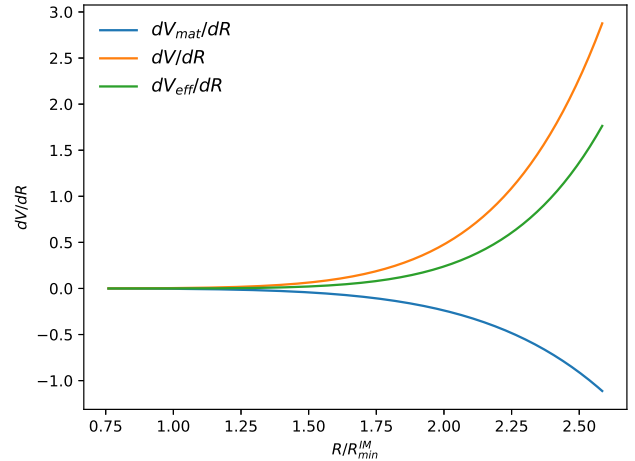}
\caption{Behavior of the quantities $dV_{\rm eff}/dR$, $dV/dR$ and $dV_{\rm mat}/dR$ in units of $10^{-76}$ {\rm cm}$^{-4}$ in the interstellar medium 
(as defined in the main text)
  with respect to $R/R_{\rm min}^{\rm IM}$ for the Starobinsky model given by Eq.~(\ref{Staromod})~\cite{Starobinsky2007} with $q=2$. A similar behavior of these quantities
  is exhibited by the other $f(R)$ models (see Secs.~\ref{sec:f(R)a} and \ref{sec:results f(R)}).}
\label{fig:dVeffdR}  
\end{figure}

According to Eq.~(\ref{R0cond}), we define the effective potential $V_{\rm eff}$ in such way that $\frac{dV_{\rm eff}}{dR}$ represents
the r.h.s of Eq.~(\ref{traceRrap}) without considering the term with $R'^2$. The latter will provide a non-linear term ${\tilde R}^2$ which
we discard in the linear approximation.
The derivative $\frac{dV_{\rm eff}}{dR}$ has two components $dV/dR:=(2f-Rf_R)/(3 f_{RR})$ and $dV_{\rm mat}/dR:=kT/(3f_{RR})$ whose behavior is depicted in
Fig.~\ref{fig:dVeffdR} taking as an example the Starobinsky model \cite{Starobinsky2007} with $q=2$ [cf. Eq.~(\ref{Staromod})].
It should be noted that for the other $f(R)$ models we scrutinize, and which produce a successful background cosmology,
the shape of the potential is very similar to  the one shown in Fig ~\ref{fig:dVeffdR}.

Although both the effective potential and its derivative are highly non-linear, the quadratic approximation of the effective potential
is acceptable as long as the effects generated by the non-linearities are confined within a
region adjacent to the two boundaries associated with the three layers (the edges of the Sun and the corona) at $r={\cal R}_\odot$ and
  $r={\cal R}_{\rm cor}$.

In order to solve Eq.~(\ref{eqRLtilde}) we impose regularity conditions at the origin $r=0$, namely ${\tilde R}_{\rm  in}^\prime(0)=0$, and the asymptotic condition
$\tilde{R}_{\rm IM}({\cal R}_{\rm IM})\approx 0$. The global solution for $R_I(r)=R^I_{\rm min}+\tilde R_I(r)$ is given by:
\begin{equation}
{\tilde R}(r)=
\begin{cases}
{\tilde R}_{\rm in} (r)= C_{\rm in}\frac{\sinh{(m_{\rm eff}^{\rm in} r)}}{m_{\rm eff}^{\rm in} r} \hspace{1cm} (0\leq r \leq {\cal R}_\odot) \\
\\
\begin{split}
{\tilde R}_{\rm cor}(r) &= C_1\frac{e^{-m_{\rm eff}^{\rm cor} r}}{m_{\rm eff}^{\rm cor} r} + C_2\frac{e^{m_{\rm eff}^{\rm cor} r}}{m_{\rm eff}^{\rm cor} r}\\
 &  \hspace{3.8cm}  ({\cal R}_\odot \leq r \leq {\cal R}_{\rm cor})\\
\end{split}
\\ \\
{\tilde R}_{\rm IM}(r)= C_{\rm IM} \frac{e^{-m_{\rm eff}^{\rm IM} r}}{m_{\rm eff}^{\rm IM} r}  \hspace{0.5cm}  ({\cal R}_{\rm cor} \leq r \leq {\cal R}_{\rm IM})
\end{cases}
\label{Rtildesol}
\end{equation}

The integration constants  $C_{\rm in}$, $C_1$, $C_2$ and 
$C_{\rm IM}$, are fixed by matching continuously the  solutions for $R(r)$ at ${\cal R}_\odot$ and ${\cal R}_{\rm cor}$. The constant $C_{\rm in}$ provides the
value of ${\tilde R}$ at the origin $r=0$. The explicit expressions for these constants are rather lengthy and not very enlightening, thus,
  we decided to omit them for brevity. When the screening effects are optimal one expects that in 
most of the Sun's interior ${\tilde R}_{\rm in} (r) \ll R_{\rm min}^{\rm in}$,
  and thus $R(r)\approx R_{\rm min}^{\rm in}$, except within a narrow region near $r={\cal R}_\odot$ where 
$R(r)$ falls-off exponentially to a value near $R_{\rm min}^{\rm cor}$. Similarly, within the corona
  $R(r)\approx R_{\rm min}^{\rm cor}$ and then again $R(r)$ decreases exponentially until $r = {\cal R}_{\rm IM}$. In this region and asymptotically ${\tilde R}_{\rm IM}({\cal R}_{\rm IM})\ll R_{\rm min}^{\rm IM}$ so that in the IM  $R_{\rm IM}(r)\approx R_{\rm min}^{\rm IM}$
  which is very small compared with the Ricci scalar near $r={\cal R}_\odot$.

At this point it is convenient to stress some differences between our solution $R(r)= R_{\rm min} + {\tilde R}(r)$
  with the solution that results from the original chameleon model $\varphi (r)$  when applied to the Sun solely
  (without the corona, for instance) and when the screening effects take place. In the
  region that extends to one A.U., and for typical values of its parameters, one finds that in the original chameleon model $m_{\rm eff}^{\rm IM} r\ll 1$, thus, the exponential term is very close to unity.
  Hence, it is the analogous of the coefficient $C_{\rm IM}$ that basically suppresses the difference $\varphi_{\rm IM} (r)-\varphi_{\rm min}^{\rm IM}$
in the solar system. Such coefficient is directly related with the famous {\it thin-shell} parameter. In fact, one can define an effective thin-shell
  parameter by including the exponential term. However, as we just emphasized, the exponential term does not contribute much in that case. Notwithstanding, in the present case, something different  occurs, namely, the exponential term, mainly in the corona and to a lesser extent in the interstellar medium,
  is fundamental to suppress ${\tilde R}(r)$ and thus, it is essential for the $f(R)$ theory to pass the solar-system tests. (It should be noted that the coefficient $C_2$ in Eq.~(\ref{Rtildesol})  is almost negligible when $m_{\rm eff}^{\rm cor}r\ge 1$ and therefore $\tilde R(r)$ in the corona behaves like a decaying exponential term).   In other words, in the    corona region, ${\tilde R}_{\rm cor}(r)$ behaves like  $|\varphi(r)-\varphi_{\rm min}|$ except that the coefficient that multiplies the Yukawa term is not necessarily ``small''.  Table~\ref{mRsol} shows that  the value of $m_{\rm eff} \cal{R}_\odot$ \footnote{Within the solar system $m_{\rm eff} r $,  differs in the more extreme case in two orders of magnitude with $m_{\rm eff} \cal{R}_\odot$ .} is different across the different regions of the solar system and this  behavior occurs in each one of the
particular $f(R)$ models analyzed in this paper.  Furthermore, it will be shown in Sec.~\ref{sec:results f(R)} that only  for those models where $m_{\rm eff}^{\rm cor} {\cal{R}_\odot} \gg 1$, the suppression is effective and the  model is able  to pass the observational bounds.  
Therefore, it is the whole combination $C_1 e^{-m_{\rm eff}^{\rm cor} r}/m_{\rm eff}^{\rm cor}$ that should be considered as root of the {\it screening} mechanism under the current approach. Also, in this scenario and with $m_{\rm eff}^{\rm IM} {\cal{R}_\odot} \ll 1$, $C_{\rm IM}/m_{\rm eff}^{\rm IM} \sim \kappa\rho_{\rm cor}{\cal{R}_{\rm cor}}$, from which it follows that the suppression of ${\tilde R}_{\rm IM}(r)$ depends mainly on the properties of the corona. This changes when $m_{\rm eff}^{\rm cor} {\cal{R}_\odot} \ll 1$ since $C_{\rm IM}/m_{\rm eff}^{\rm IM} \sim \kappa (\rho_{\rm cor}+\rho_{\odot}){\cal{R}_{\odot}}$ causing the suppression to be much less effective.

\begin{table}[!hbtp]
\centering
\begin{tabular}{|l|l|l|l|}
\hline
\textbf{$f(R)$}   &  \textbf{Sun} &  \textbf{Corona} &  \textbf{\rm IM}    \\ \hline
Starobinsky $q=2$ & $2.80\times10^{69}$ & $9.60 \times 10^{23}$ & $9.54 \times 10^{-4}$ \\ \hline
Starobinsky $q=0.4$ & $1.39\times10^{23}$ & $8.42 \times 10^{1}$ & $2.12 \times 10^{-11}$ \\ \hline
Hu-Sawicki $n=4$ & $6.53\times10^{69}$ & $2.23 \times 10^{24}$ & $2.23 \times 10^{-3}$ \\ \hline
MJWQ & $1.10\times10^{12}$ & $7.66 \times 10^{-4}$ & $7.66 \times 10^{-13}$ \\ \hline
\end{tabular}
\caption{The dimensionless quantity $m_{\rm eff} \cal{R}_\odot$ 
evaluated in the three media proposed for our analysis (columns 2--3) assuming the four non-linear $f(R)$ models 
described in Sec.~\ref{sec:f(R)a}.}
\label{mRsol}
\end{table}

Using the solution (\ref{Rtildesol}) together with (\ref{rho}) in Eq.(\ref{psiprimesun4}) and assuming
the conditions $f_{R}^{\rm min,I}\sim 1$, ${R}_{\rm min}^I \sim f({R}_{\rm min}^I)$, $r^2 {R}_{\rm min}^I \ll 1$ with 
${R}_{\rm min}^I$ satisfying Eq.~(\ref{minconpot}) in the three regions and after integration, we obtain for the corona and the IM the following solutions:

\begin{eqnarray}
 \psi_{\rm cor}(r)&=&\frac{\tilde{R}_{\rm cor}(r)}{6 m_{\rm eff}^{\rm cor}}\left(r-\frac{1}{m_{\rm eff}^{\rm cor}}\right)-2 C_1 \frac{e^{-m_{\rm eff}^{\rm cor} r}}{6 m_{\rm eff}^{\rm cor}} \nonumber\\
&& + \frac{C_\psi^{\rm cor}}{r} \;, \\  
 \psi_{\rm IM}(r)&=&\frac{- \tilde{R}_{\rm IM}(r)}{6} \frac{1+m_{\rm eff}^{\rm IM} r }{(m_{\rm eff}^{\rm IM})^2}+\frac{C_\psi^{\rm IM}}{r}\;,
\end{eqnarray}
where we kept only the leading terms in agreement with the qualitative arguments presented in the previous section. 

Here $C_\psi^{\rm cor}$ and $C_\psi^{\rm IM}$ are integration constants that are fixed when matching both solutions at ${\cal R}_{\rm cor}$ and
when matching $\psi_{\rm cor}(r)$ with the interior solution $\psi_{\rm in}(r)$ at ${\cal R}_\odot$. Keeping the leading terms, we find: 

\begin{eqnarray}
\begin{split}
C_\psi^{\rm cor}=&\frac{{\cal R}_\odot^3}{12 f_{R_{\rm min}^{\rm in}}} (f_{\rm min}^{\rm in}+\kappa \rho_\odot)+\frac{{\cal R}_\odot C_1 e^{-m_{\rm eff}^{\rm cor} {\cal R}_\odot}}{3 m_{\rm eff}^{\rm cor}}\\
 &  -\frac{\tilde{R}_{\rm cor}({\cal R}_\odot){\cal R}_\odot}{6 m_{\rm eff}^{\rm cor}}\left({\cal R}_\odot-\frac{1}{m_{\rm eff}^{\rm cor}}\right)\;, \\
\end{split}\\
C_\psi^{\rm IM}=C_\psi^{\rm cor}+{\cal R}_{\rm cor} \tilde{R}_{\rm IM}({\cal R}_{\rm cor})\frac{1+m_{\rm eff}^{\rm IM} {\cal R}_{\rm cor}}{6 \left(m_{\rm eff}^{\rm IM}\right)^2}\;,
\end{eqnarray}
when assuming $m_{\rm eff}^{\rm IM} \ll m_{\rm eff}^{\rm cor}\ll m_{\rm eff}^{\rm in}$, $ R_{\rm min}^{\rm IM} \ll 
R_{\rm min}^{\rm cor} \ll R_{\rm min}^{\rm in}$ and $m_{\rm eff}^{\rm in} {\cal R}_\odot\gg 1$ we obtain:

\begin{eqnarray}
    C_1 &=& e^{m_{\rm eff}^{\rm cor}({\cal R}_{\rm cor}+{\cal R}_\odot)} \left[\frac{e^{m_{\rm eff}^{\rm cor}{\cal R}_\odot}  R_{\rm min}^{\rm cor} (1+m_{\rm eff}^{\rm IM} {\cal R}_{\rm cor})}{ m_{\rm eff}^{\rm cor}(e^{2 m_{\rm eff}^{\rm cor}{\cal R}_\odot}+e^{2 m_{\rm eff}^{\rm cor}{\cal R}_{\rm cor}})} \right.\nonumber \\
&&    \left.+  \frac{e^{m_{\rm eff}^{\rm cor}{\cal R}_{\rm cor}}  R_{\rm min}^{\rm in}  {\cal R}_\odot}{ e^{2 m_{\rm eff}^{\rm cor}{\cal R}_\odot}+e^{2 m_{\rm eff}^{\rm cor}{\cal R}_{\rm cor}}}\right].
\end{eqnarray}

Regarding Eq.~(\ref{psimphi}), we can integrate it in the corona region and in the IM using the solution (\ref{Rtildesol}) 
and neglecting the pressure term $T^{r}_{\,\,r}$ and taking $T^{t}_{\,\,t}=-\rho_{\rm cor,IM}$ to obtain: 
 \begin{eqnarray}
  \phi_{\rm cor}(r)&=& \psi_{\rm cor}(r)+\tilde{R}_{\rm IM}({\cal R}_{\rm cor}) \frac{f_{RR_{\rm min}^{\rm IM}}}{f_{R_{\rm min}^{\rm IM}}}
+\frac{\tilde{R}_{\rm IM}({\cal R}_{\rm cor}) {\cal R}_{\rm cor}}{6 m_{\rm eff}^{\rm IM}} \nonumber \\
 & & +\frac{ \kappa \rho_{\rm cor}}{4 f_{{R}_{\rm min}^{\rm cor}}}({\cal R}_{\rm cor}^2-r^2)+ \tilde{R}_{\rm cor}(r) \frac{f_{RR_{\rm min}^{\rm cor}}}{f_{R_{\rm min}^{\rm cor}}} \;, \label{phioutcor}\\
  \phi_{\rm IM}(r) &=& \psi_{\rm IM}(r)-\frac{r^2 \kappa \rho_{\rm IM}}{4 f_{R_{\rm min}^{\rm IM}}}+\tilde{R}_{\rm IM}(r) \frac{f_{RR_{\rm min}^{\rm IM}}}{f_{R_{\rm min}^{\rm IM}}} 
+\frac{\tilde{R}_{\rm IM}(r) r}{6 m_{\rm eff}^{\rm IM}}.
\nonumber \\
  \label{phiout}
\end{eqnarray}
The above equation has an additional term proportional to $\tilde{R}_{\rm IM}(r) r$ 
when compared with Eq.~(\ref{psimphiout}). As we emphasized before, that term is also very small when the screening ensues, thus it is not very important if one includes it or excludes it ultimately. 
In Eqs.(\ref{phioutcor}) and  (\ref{phiout}) we also keep the terms proportional to $r^2  \rho_{\rm cor,IM}$ although they 
are negligible in the corona and the IM regions.

As can be noted from Eqs.~(\ref{phioutcor}) and  (\ref{phiout}), the perturbations $\phi_I(r)$ have the form $\phi_{\rm cor, IM}(r)=\psi_{\rm cor,IM}(r) + \alpha_{\rm cor,IM}(r)$, where $\alpha_{\rm cor,IM}(r)$ can be read off from those equations.
Finally, the PNP $\gamma$ is defined as $\gamma(r)=\frac{\psi(r)}{\phi(r)}$, or equivalently  $\gamma(r)=\frac{1}{1+\epsilon(r)}$ where

\begin{equation}
\epsilon (r)=
\begin{cases}
\frac{\alpha_{cor}(r)}{\psi_{cor}(r)} \hspace{0.8cm} ({\cal R}_\odot\leq r \leq {\cal R}_{\rm cor})\\ \\
\frac{\alpha_{\rm IM}(r)}{\psi_{\rm IM}(r)}  \hspace{0.8cm}  ({\cal R}_{\rm cor} \leq r \leq {\cal R}_{\rm IM})
\end{cases}
\end{equation}
We see then, that the PNP $\gamma$ as defined above, depends actually on the coordinate $r$ and on the $f(R)$ model.
  Thus, we have to ensure that $\gamma(r)$ outside
  the Sun (i.e., in the corona and in all the solar-system neighborhood) has to satisfy the observational bounds (\ref{gamabnd})
  for the cosmologically viable $f(R)$ models.

\section{Results for different $f(R)$ models}
\label{sec:results f(R)}

We have applied the approach described in Section \ref{IV} to obtain a prediction for the PNP $\gamma$ (through an estimate for $\epsilon$) for four $f(R)$
specific models that have been presented in Sec.\ref{sec:f(R)a}. 
  
The Cassini mission established the constraint (\ref{gamabnd}) by measuring the Shapiro time delay of
radio signals~\cite{Berlotti2003}, and the VLBI interferometer also put a constraint $|\gamma-1|<4 \times 10^{-4}$ measuring the light deflection due to the Sun \cite{vlbi,Shapiro2004}.

For the parameter space analyzed in each $f(R)$ model, the linearization of the effective potential inside the Sun 
turns to be a good approximation. For instance, the condition Eq.(\ref{TSP}) is satisfied for all the $f(R)$ models described below. The Compton condition
($L\gg (m_{\rm eff}^{\rm cor})^{-1}$) for the corona is also satisfied for all except for the logarithmic MJWQ model.

\begin{enumerate}
 
\item \textbf{The Starobinsky $f(R)$ model}

When $q \geq 2$ we find that $\tilde{R}_{\rm cor}(r)$ and $\tilde{R}_{\rm IM}(r)$ are negligible almost everywhere as can be seen from Fig.~\ref{fig:Staro2R} for the particular case $q=2$. 
The largest gradients of $R(r)$ are confined within a narrow shell near the surface of the Sun and adjacent to the corona (see Fig.~\ref{fig:Staro2Rzoom} for a zoom).\\
As we increase $q$, the effective mass increases with $q$ as well since
\begin{eqnarray}
  m_{\rm eff}^2 &=&\frac{-2(1+q)R_{\rm min}^3}{3[(1+2q)R_{\rm min}^2-R_s^2]}\nonumber\\
  &+& \frac{(R_{\rm min}^2+R_s^2)^2(1+\frac{R_{\rm min}^2}{R_s^2})^q}{6\lambda q R_s[(1+2q)R_{\rm min}^2-R_s^2]}\;, 
\end{eqnarray}
causing $e^{-m_{\rm eff}^{\rm IM }r}/r \ll e^{-m_{\rm eff}^{\rm cor }r}/r \ll 1$ and making
the regions where ${\tilde R}_{\rm cor, IM}$ is confined
even narrower and closer to the Sun's border.
Thus, $R(r)$ behaves basically like a {\it single} step function
(similar to GR under the constant density model for the Sun's interior) given that, outside the Sun ${\cal R}_\odot\lesssim r$,
the Ricci scalar is already very small compared with $R_{\rm min}^{\rm in}\sim \kappa \rho_{\odot}$ inside the Sun. This behavior implies that the PNP $\gamma$ satisfies the experimental bounds within the solar system by several orders of magnitude (see Fig.\ref{fig:Staro2gamma}).
From Figure~\ref{fig:denscor} one appreciates that decreasing the density of the corona
  reduces the {\it screening} effect as it generates smaller gradients on $R(r)$. Thus, if one does not include the corona,
  which is equivalent to reduce the corona's density to the IM value, the screening effects become less effective.
  This behavior is common in the other $f(R)$ models discussed below.

\begin{figure}[t]
\includegraphics[width=0.45\textwidth,right,trim={3cm 8cm 0.75cm 8cm}]{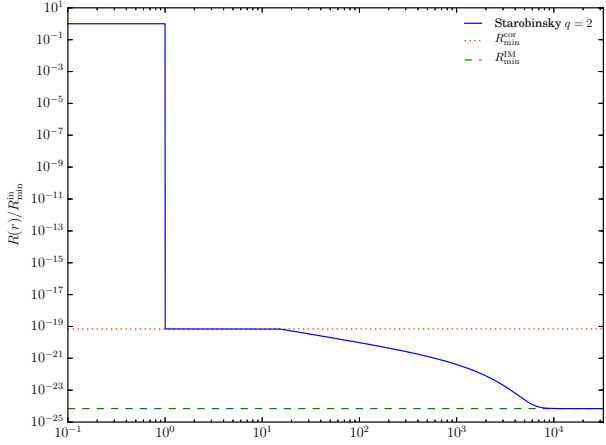}
\caption{Ricci scalar (in units of $R_{\rm min}^{\rm in}$) as a function of $r$ (in units of ${\cal R}_\odot$) computed from the Starobinsky model with $q=2$. Inside the Sun $R(r) \approx R_{\rm min}^{\rm in}$ and beyond ${\cal R}_\odot$ the Ricci scalar $R(r)\ll R_{\rm min}^{\rm in}$. The total solution behaves basically as a single step function. The red dotted line and the green dashed line indicate the values $R_{\rm min}^{\rm cor}$ and $R_{\rm min}^{\rm IM}$, respectively (both values are very small compared with $R_{\rm min}^{\rm in}$).}
\label{fig:Staro2R}
\end{figure}

\begin{figure}[t]
\centering
\includegraphics[width=0.58\textwidth,trim={3cm 8cm 0.5cm 8cm}]{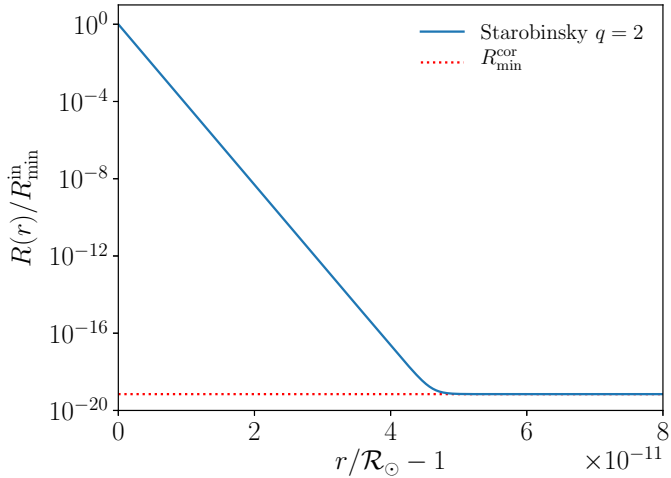}
\caption{Same as Figure~\ref{fig:Staro2R}, but plotted very near the Sun's surface within the corona.  
The Ricci scalar decreases around 19 orders of magnitude from $R_{\rm min}^{\rm in}$ to $R_{\rm min}^{\rm cor}$ 
within a very thin region $\Delta {\cal R}_\odot\sim 5\times 10^{-11}{\cal R}_\odot$ (between $r={\cal R}_\odot$ and 
$r= {\cal R}_\odot + \Delta {\cal R}_\odot$).}
\label{fig:Staro2Rzoom}  
\end{figure}

\begin{figure}[t]
\centering
\includegraphics[width=0.6\textwidth,trim={3cm 8cm 0.5cm 8cm}]{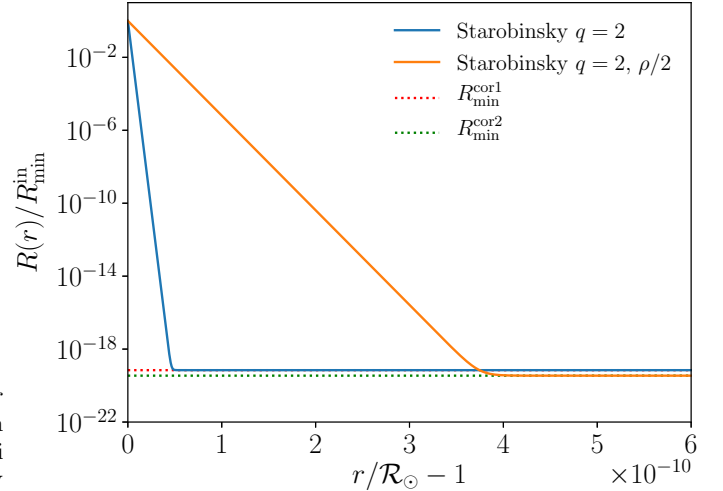}
\caption{The figure depicts the effect on $R(r)$ when decreasing the corona's density by half.
    The higher the density the steeper the gradient of $R(r)$ provoking the {\it screening} effect to be more effective.}
\label{fig:denscor}  
\end{figure}

\begin{figure}[t]
\centering
\includegraphics[width=0.49\textwidth, trim={1cm 6.5cm 2cm 6.5cm}]{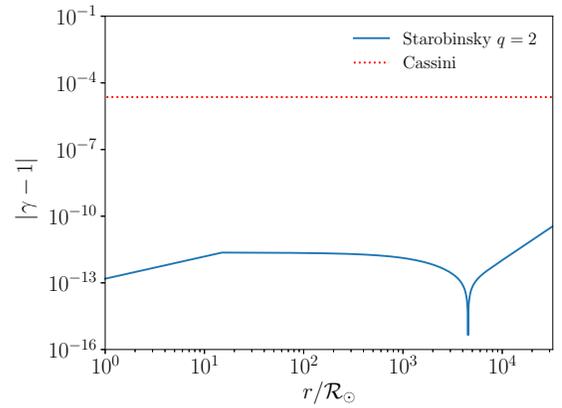}
\caption{Deviation parameter $|\gamma-1|$ as a function of $r$ (in units of ${\cal R}_\odot$) computed from the Starobinsky model 
used in Figure~\ref{fig:Staro2R}. The constraint from the Cassini mission is shown as a horizontal dotted red line.}
\label{fig:Staro2gamma}  
\end{figure}

On the other hand, taking $q=0.4$ and $\lambda=4$, which satisfy the constraints considered in Sec. II of Starobinsky's paper\cite{Starobinsky2007}, 
the Ricci scalar $R(r)$ differs substantially from the case $q=2$ and $\lambda=1$ (cf. Fig.~\ref{Staro04R}). As a consequence the parameter
$\gamma$, as depicted by Fig.~\ref{Staro04gamma}, fails the solar-system tests by more than four orders of magnitude.
The difference between both cases can be explained by noting that when $q=2$, $m_{\rm eff}^{\rm cor} r \gg 1$ while
if $q=0.4$,  $m_{\rm eff}^{\rm cor} r \ll 1$ and therefore the Yukawa term in the corona produces much more screening
  as $q$ is increased.

\begin{figure}[t]
\centering
\includegraphics[width=0.49\textwidth,,trim={1cm 6.5cm 2cm 6.5cm}]{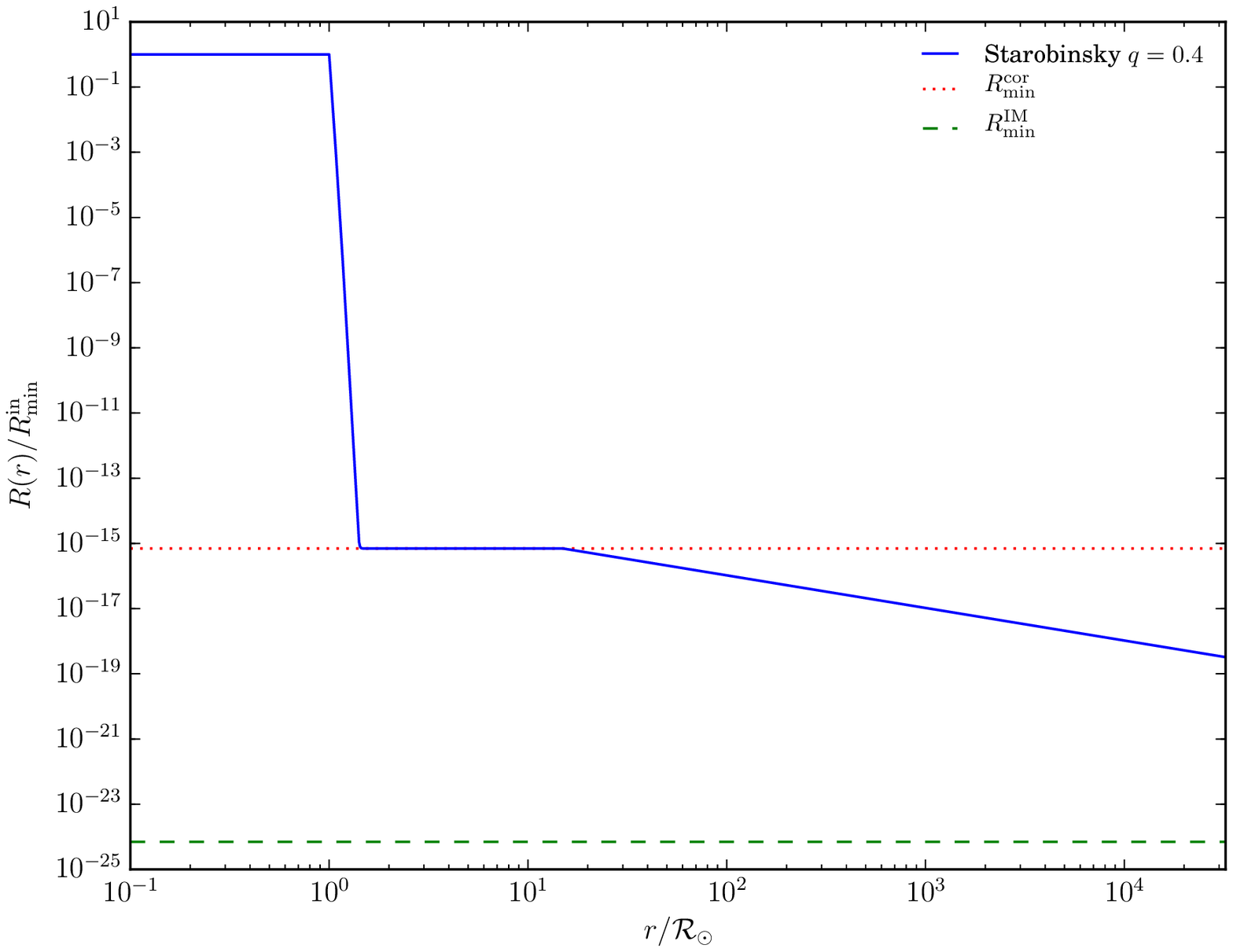}
\caption{Similar to Figure~\ref{fig:Staro2R} but taking $q=0.4$ and $\lambda=4.0$ for the Starobinsky model.
    Qualitatively, the Ricci scalar behaves similarly in both cases ($q=2$ and $q=0.4$), however, for $q=0.4$ it is several orders of magnitude larger
    than the case $q=2$, and thus, this second model fails the solar-system tests.}
\label{Staro04R}  
\end{figure}

\begin{figure}[t]
\centering
\includegraphics[width=0.49\textwidth,trim={1cm 6.5cm 2cm 6.5cm}]{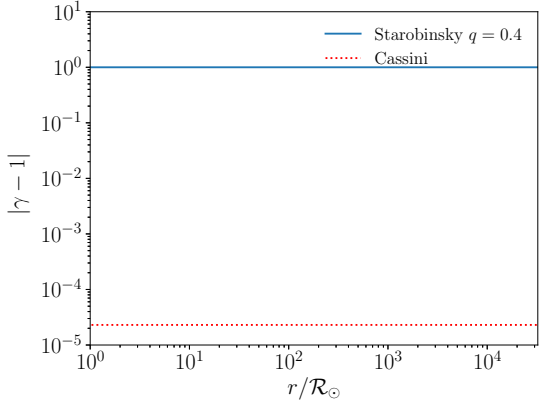}
\caption{Similar to Figure~\ref{fig:Staro2gamma} using $q=0.4$ and $\lambda=4.0$. This Starobinsky model fails the solar-system tests: 
the predicted value for $|\gamma-1|$ is around four orders of magnitude larger than the observational bounds.}
\label{Staro04gamma}  
\end{figure}

 \item \textbf{The Hu-Sawicki $f(R)$ model}

For $n \geq 4$, $R(r)$ has basically the same behavior that the Starobinsky model has with $q \geq 2$ since $m_{\rm eff}$ increases with $n$
in a similar way as the effective mass of the previous model does with $q$. We conclude that the observational bounds 
imposed on $\gamma$ are satisfied for this model as well.

\item \textbf{Exponential $f(R)$ model}

For this model with $R_{*}=2.5H_0^2$ and $\beta=2$, the behavior of $R(r)$ is practically the same as the successful Starobinsky and Hu-Sawicki models, thus, the solar-system tests are passed
also in this case. 

\item \textbf{MJWQ $f(R)$ model}

 Figure~\ref{MirandaR} shows that the Ricci scalar $R(r)$ with $R_{\rm m}=H_0^2$ and $\alpha=2$ does not reach its minimum in the corona, not even in the IM region, 
 and decreases very slowly with a profile different from a step function. In particular, this behavior 
is because the Compton condition ($L\gg (m_{\rm eff}^{\rm cor})^{-1}$) is not satisfied in the corona. Therefore, $m_{\rm eff}^{\rm cor}$ is replaced by $L^{-1}$ and the approximation of the effective potential is performed around a  ``new'' $R_{\rm min}^{\rm cor}$ which is not anymore a minimum of the effective potential and satisfies $L^{-2}=(f_R-Rf_{RR})/(3f_{RR})|_{R_{\rm min}^{\rm cor}}$ \cite{Burrage15,Schogel16}. However, overall, the method is at least almost the same and the result for $R(r)$ can be seen in Fig.(\ref{MirandaR}). 
In this case $\tilde{R}(r)$ is not suppressed at the exterior of the Sun and as a consequence the parameter $\gamma$ is 
incompatible with the observations (cf. Fig.~\ref{Mirandagamma}).

\begin{figure}[t]
\centering
\includegraphics[width=0.49\textwidth,trim={1cm 6.5cm 2cm 6.5cm}]{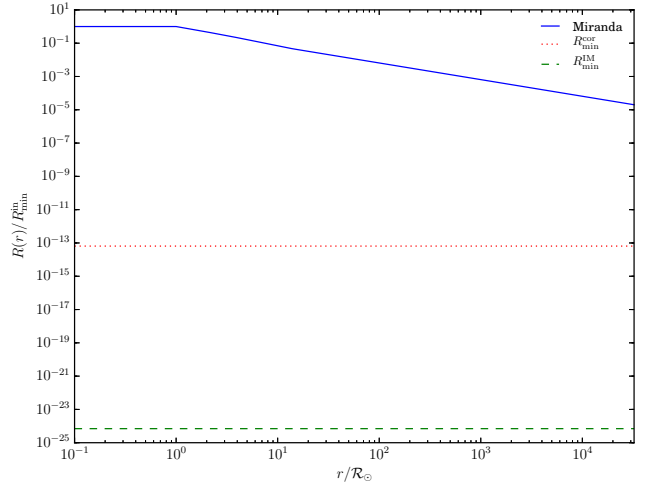}
\caption{Similar to Figure~\ref{fig:Staro2R}, but for the MJWQ model. 
Inside the Sun $R(r) \approx R_{\rm min}^{\rm in}$, but outside, the Ricci scalar 
does not reach a minimum $R_{\rm min}^{\rm IM}$ and decreases slowly.}
\label{MirandaR}  
\end{figure}

\begin{figure}[t]
\centering
\includegraphics[width=0.49\textwidth,trim={1cm 6.5cm 2cm 6.5cm}]{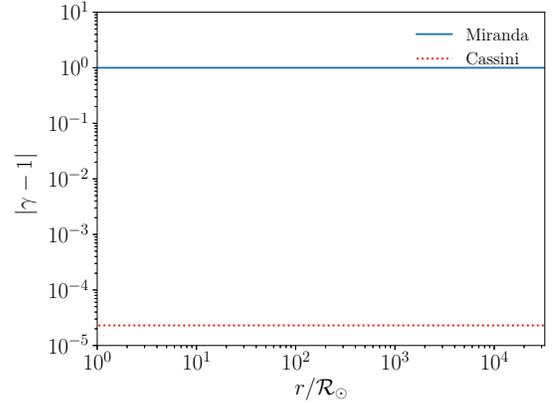}
\caption{Similar to Figure~\ref{fig:Staro2gamma}, but for the MJWQ model. 
This model does not satisfy the observational bounds on $\gamma$ by more than four orders of magnitude.}
\label{Mirandagamma}  
\end{figure}
\end{enumerate}

\section{Comparison with other results}
\label{sec:review}

In this section we compare our analysis with other studies from the past~\cite{Hu2007,Faulkner2007,Chiba2007,Sotiriou2010,Capozziello08,Guo2014} 
where $f(R)$ models were confronted with the solar system, as well as with other tests, and where
the PNP $\gamma$ is estimated following different approaches and perturbative techniques.

In the analysis by Hu \& Sawicki~\cite{Hu2007}, a linear approximation for the metric perturbations is performed using isotropic coordinates, 
as opposed to the area coordinates used in the present work. The fundamental field used there for the chameleon-like analysis is $f_R^{HS}$\footnote{ 
According to the notation used here $f_{\rm ours}(R)= R+ f^{HS}(R)$.} instead of $R$ itself. In that work the authors 
assume $|f_R^{HS}|\ll 1$ and $f^{HS}/R\ll 1$ {\it ab initio} and proceed by considering only the leading terms. They use 
the quantity $\Delta f_R^{HS}:= f_R^{HS}-f_{R\infty}^{HS}$ as a perturbation around an approximate Minkowski background, where $f_{R\infty}^{HS}$ is the value associated with the ``galaxy'' 
($f_{R\infty}^{HS}= f_{R}^{HS} (R_g)$ where $R_g\approx \kappa \rho_g$ with $\rho_g= 10^{-24}{\rm g\,cm^{-3}}$), 
which is the equivalent of our $\rho_{\rm IM}$. They consider an inhomogeneous model for the Sun and the corona, and thus, the density 
is not a simple step function. Under those approximations the quantity $-\Delta f_R^{HS}$ turns out to be exactly equal to one 
of the metric perturbations, and thus, is directly related with the $\gamma$ parameter ({\it mutatis mutandis} $|\Delta f_R^{HS}|$ would 
correspond to our $|\psi-\phi|$, and thus $|\gamma_{HS}-1|= |\Delta f_R^{HS}/\phi|= |1-\psi/\phi|$). Hence, as far as $\Delta f_R^{HS}\ll 1$ 
one recovers the observational bounds on $\gamma_{HS}$. The cosmological value $f_{R0}^{HS}$ used by HS is such that 
$(f_{Rg}^{HS}/f_{R0}^{HS})^{\frac{1}{n+1}}\sim 8\times 10^{-7} R_0/m^2$. However, their results on $|\gamma_{HS}-1|\sim 10^{-15}$ 
for $n=4$ are almost insensitive to the possible values $f_{R0}^{HS}$ in the range $0$--$0.1$. The Compton wave-length is 
$\lambda_{HS}\sim 8300 {\cal R}_\odot |f_{R0}^{HS}|$, which is much smaller than the solar-system size used there $r\sim 10^6{\cal R}_\odot $. 
Therefore, the Yukawa factor $e^{-r/\lambda_{HS}}/r\ll 1$ suppresses the perturbed field $\Delta f_R^{HS}$ considerably in the solar-system 
neighborhood.

Faulkner {\it et al}~\cite{Faulkner2007} also use the variable $\varphi=f_R$ as the fundamental chameleon field but perform all their calculations 
in the Einstein frame and after a long chain of steps they return to the Jordan frame in order to compare their results with the bounds on $\gamma$. 
They also use area $r$ coordinates  as we do. For the model $f_F(R)= R + (m-1) \mu^2 (R/\mu^2)^m-2\Lambda$ considered by them,
the Compton wavelength condition $10^{10}$ A.U.$\lesssim \lambda_{F}$ is found, and thus, the Yukawa factor does not contribute much to suppress the field, but it is rather their thin shell parameter 
$\Delta$ which is responsible for the suppression, provided the parameter $\mu^2/H_0^2$ satisfies some constraints that depend on the exponent $m\neq 1$. 
At first sight it is intriguing that the authors include an explicit cosmological constant $\Lambda$ since precisely one of the goals of $f(R)$ gravity is to produce dynamically an effective 
$\Lambda_{\rm eff}$ without an explicit $\Lambda$. However, it is after the authors consider the case $m\ll 1$, for which their model reduces to a logarithmic 
one, that one realizes that the condition $\mu^2/H_0^2\sim 10^{-6}\ll 1$, is required to pass the solar-system tests, and therefore, that  the only possibility 
to recover the cosmological observations as well is when $\Lambda\neq 0$, in which case their model becomes almost indistinguishable from $f_{GR}(R)$. 
In other words, it seems that their $f_F(R)$ model with $\Lambda=0$ would be 
ruled out whether by the solar system or by the cosmological observations, in particular if $m\ll 1$. The problem with this model is similar but opposite to 
the logarithmic model MJWQ (\ref{fMJWQ}), which can produce a relatively adequate background cosmology without an explicit $\Lambda$, but fails 
the solar-system tests. Notice that in the MJWQ the coefficient $\alpha R_{\rm m}\sim H_0^2$, whereas in the Faulkner {\it et al.} with $m\ll 1$, 
the coefficient is very small ($\mu^2\sim 10^{-6} H_0^2$). Conversely, if in the MJWQ one takes $\alpha R_{\rm m}\ll H_0^2$, the solar-system tests would be 
recovered, but not the cosmological ones.

Another interesting analysis was performed by Guo~\cite{Guo2014}, which like in \cite{Hu2007,Faulkner2007}, the author promotes the variable $f_R$ as 
the fundamental field, and uses area coordinates as we do. The author remains in the Jordan frame as well. 
After attempting a first naive analysis, which is similar to ours as presented in Appendix~\ref{sec:wronglin}, leading to $\gamma=1/2$, 
Guo implements the necessary modifications to recover the screening effects, only to realize that for an $f(R)$ model to be compatible with 
the solar system observations the function $A(R)= f(R)-R$ and its derivatives have to be very small compared with the GR expectations, namely 
$|A|\ll R$, $|f_R|\ll 1$ and $|f_{RR} R|\ll 1$, which is somehow the same conclusion found in \cite{Faulkner2007}. One faces again the same
dilemma discussed above, that is, at cosmological scales $f(R)$ gravity has to produce a late accelerated expansion but with a dynamic geometric dark energy,
while, at local scales, the theory has to respect the solar-system tests. 
Thus, one can achieve very easily the above conditions in the solar system, but 
by failing the cosmological observations, unless an explicit $\Lambda$ is introduced, in which case, one simply returns to the argument we discussed above 
within the Faulkner {\it et al} model. The challenge consists of precisely  manufacturing {\it non linear} $f(R)$ models that satisfy all the possible tests. 
For instance, clearly at the cosmological level one requires $|A|\sim R$ in order to produce an adequate cosmic acceleration, while in the solar system 
neighborhood $|A|\ll R$. Thus, in both scenarios the ``dynamics'' of the model should be responsible to achieve what is needed to be successful, otherwise
$f(R)$ theory without an explicit  $\Lambda$ is basically an end road\footnote{In our physics community sometimes authors take a different 
point of view and consider any non-linear $f(R)$ gravity (including $\Lambda$), or any other gravitational theory for that matter, 
and then look to all its possible predictions, but without having any specific goal, like explaining a yet unexplained phenomenon. 
There will be some of those theories which will be compatible with the current observations while predicting 
new effects that could be validate or ruled out by new (unperformed) experiments. In other words, some physicist embrace the lemma 
by T. H. White, {\it what is not forbidden is compulsory} and then analyze the consequences of the proposed theory. Here, however, we have a very specific 
goal, which is to produce a {\it geometrical} dark energy model compatible with all the possible observations.}. In order to understand better the 
chameleon mechanism, Guo proposes to use quantum-tunneling or instanton analogues, but at the end the author is compelled to solve numerically 
a non-linear equation for the field $f_R$  as everybody else. Guo analyzes a logarithmic $f(R)$ model and also the simplest HS model ($n=1$) 
using a non-homogeneous density model for the Sun immersed within a constant density background. The author concludes that the logarithmic model is basically ruled out. As for the HS model, the author solves a kind of chameleon equation, and from its behavior, claims that the solar-system tests are passed but without 
offering a further scrutiny on $\gamma$ as we do here. Finally, we stress that the review article~\cite{Sotiriou2010}, 
which follows closely the one by Chiba~\cite{Chiba2007} as regards the discussion on the the solar-system tests, only provides the naive approximation that neglects the screening effects 
(see Appendix \ref{sec:wronglin}) which is unsuitable to recover $\gamma\sim 1$. The authors do not perform
explicitly the analysis by taking into account the screening effects, but mention briefly some possible solutions using it.

\section{Conclusions}
\label{sec:conclusions}
In this paper we have performed a thorough and careful, albeit simplified, systematic analysis of $f(R)$ models within the framework of the solar-system tests 
in the Jordan frame. We use the Ricci scalar itself as fundamental variable and provide the full system of equations required to perform both, the full non-linear 
as well as a non-standard perturbative analysis. We limit ourselves to the latter and leave the full non-linear study for the future. The non-standard linear 
analysis is similar to the one performed in the original chameleon equation, which consists of perturbing the field around the minima that 
appear when considering the effective potentials in several media of different densities. In the present case, we consider three media, the Sun, its corona, 
and the IM. Thus, we solve a kind of perturbed chameleon equation for $R$ in the three media, and match the solutions at the boundary layers. 
The matching of these solutions is the mechanism that allows one to incorporate the non-linearities in the model in a simplified way. As far as we are aware, 
this is the first time that this kind of approach has been applied to $f(R)$ models directly without the incorporation of new field variables that 
can be problematic (i.e. leading to multivalued potentials). A similar approach is followed by Guo~\cite{Guo2014}, but ultimately the author is 
unsuccessful in providing a clear-cut and general method, and at the end the analysis is fairly inconclusive in the relationship between 
the screening mechanism in $f(R)$ and the $r$ dependence of the effective $\gamma$ parameter. On the other hand, the Hu-Sawicki 
perturbative analysis is rather exhaustive but limits itself to their model without providing the full non-linear equations for the 
static and the spherically symmetric case. Other studies~\cite{Brax2008}, simply follow, {\it mutatis mutandis}, the chameleon-like methods 
in the Einstein frame that we discussed in our comparison Section~\ref{sec:review} with the limitations they entail when applied to $f(R)$ gravity. 

We hope to overcome our numerical limitations and implement better techniques in order to tackle the full non-linear problem which consists of solving 
Eqs.~(8)--(10), in the solar system neighborhood, even if the gravitational field there is weak. That study will allow us to quantify more clearly the extent 
to which our non-standard linear analysis is reliable. Notably, when the sources become more compact and gravity stronger.


\section*{Acknowledgments}
The authors acknowledge the use of the superclusterMIZTLI of UNAM through project LANCAD-UNAM-DGTIC-132 and thank the people of DGTIC-UNAM for technical and computational support.

MS was supported in part by DGAPA--UNAM grants IN107113, IN111719, and SEP--CONACYT grant CB--166656. CN, LK. and SL. are supported  by the National Agency for the Promotion of Science and Technology (ANPCYT) of Argentina grant PICT-2016-0081; and grant G140 from UNLP.
\appendix

\section{The naive (incorrect) linear analysis}
\label{sec:wronglin}

For pedagogical purposes, and for completeness, we provide the incorrect naive analysis of Eq.~(\ref{traceRrap}) that led to the
  spurious conclusion on the unviability of all the non-linear $f(R)$ models within the solar system. The analysis consists
  of linearizing Eq.~(\ref{traceRrap}) around one minimum only  which is taken to be the cosmological value
  $R_0$. This value provides a non vanishing effective cosmological constant
  $\Lambda_{\rm eff}= R_0/4$ which is responsible for explaining the late acceleration expansion of the Universe within the framework of
  $f(R)$ gravity. That is, one proceeds with a {\it standard} perturbation scheme with a single background for $R(r)$ which is given only by $R_0$. This incorrect analysis assumes implicitly that the scalar DOF perturbation ${\tilde R}_{\rm out}(r)$
  is not suppressed whatsoever outside the Sun by neglecting all possible non-linearities of the chameleon type.
  Furthermore, this analysis also supposes, as one usually does in GR,
  that outside the Sun there is a vacuum. Thus, the contribution ${\tilde R}_{\rm out}(r)$ to the metric perturbations  leads to order one deviations
  on the PNP $\gamma$. While this analysis is well known in the literature~\cite{Chiba2007,Sotiriou2010,Oyaizuetal,Guo2014}
leading to a value $\gamma\approx 1/2$ inconsistent with the observations, not all the analyses
are exactly the same, although, equivalent. Thus, it is enlightening to recover the same (wrong) conclusion from the current formalism
and in spherical symmetry using the Ricci scalar as the scalar DOF instead of the variable $\chi$
  \footnote{In most, if not all, of the approaches presented in the literature, they consider second order equations for the metric
    perturbations.}.

Let us consider 
Eq.~(\ref{traceRrap}) and linearize it around the value $R_0$ under condition Eq.~(\ref{R0cond}) but in a  vacuum. We obtain
\begin{equation}
\label{traceRraptilde}
{\tilde R}'' + \frac{2 {\tilde R}'}{r} \approx \frac{\kappa T}{3f_{RR_0}} + m_{{\rm eff},0}^2 {\tilde R}  \;,
\end{equation}
where the effective mass is taken like in (\ref{meff1}) but in vacuum and evaluated at the minimum $R_0$:
\begin{equation}
\label{meff}
m_{{\rm eff},0}^2= \frac{2f_0- R_0^2f_{RR_0}}{3 R_0 f_{RR_0}} \;,
\end{equation}
and the minimum $R_0$ satisfies Eq.(\ref{R0cond}) taking $T\equiv 0$:
\begin{equation}
\label{R0vac}
\frac{2f - R f_{R}}{3f_{RR}} \Big{|}_{R_0}=0 \;.
\end{equation}
In (\ref{traceRraptilde}) the contribution $T$ for the matter perturbation is taken only within the Sun. Outside $T\equiv 0$.
Thus, this is a generic equation, and below we provide its solutions inside and outside the Sun and their matching
at its surface.

From now on, we assume that $f(R)$ is a non-linear model, and thus
$f_{RR_0}\neq 0$. Furthermore, we require $f_{RR_0}>0$ in order to avoid tachyonic instabilities~\cite{Dolgov2003}. This condition holds in all viable cosmological nonlinear $f(R)$ models analyzed so far. Since $R_0\sim H_0^2$, then
$m_{{\rm eff},0} r\ll 1$ within the solar system, and Eq.~(\ref{traceRraptilde}) reduces even further:
\begin{equation}
\label{traceRraptildeap}
{\tilde R}'' + \frac{2 {\tilde R}'}{r} \approx \frac{\kappa T}{3f_{RR_0}}  \;.
\end{equation}
Moreover, at $R_0$ we have
\begin{equation}
\label{R0vac2}
2f_0 - R_0 f_{R_0}=0 \;.
\end{equation}

At this point we cannot take the GR limit any longer, since Eq.(\ref{traceRraptildeap}) implies that the relationship between ${\tilde R}$ and $T$ is differential and not algebraic. In principle, the recovering of the GR expectations at the solar system would come naturally from the behavior of ${\tilde R}$,
but this will not be the case given that we have inhibited implicitly the possibility of any kind of suppression by a {\it screening}
mechanism by virtue of the wrong assumptions.

By using the non-relativistic and incompressible fluid approximation $T\approx -\rho_\odot= const$ in the interior of the Sun, and neglecting the IM, we can solve Eq.(\ref{traceRraptildeap}) inside and outside the Sun and match the two solutions continuously at ${\cal R}_\odot$. The final result is

\begin{equation}
\label{solRtildew}
{\tilde R}(r)=
\begin{cases}
{\tilde R}_{\rm in} (r)= \frac{\kappa M_\odot}{8\pi  {\cal R}_\odot f_{RR_0}}\left(1-\frac{r^2}{3 {{\cal R}_\odot}^2} \right) \hspace{0.3cm} (0\leq r \leq {\cal R}_\odot) \\
 \\
{\tilde R}_{\rm out}(r)=  \frac{\kappa M_\odot}{12\pi f_{RR_0}} \frac{1}{r} \hspace{1cm}  ({\cal R}_\odot \leq r \lesssim 150 \, {\rm A.U.})
\end{cases}
\end{equation}
where we imposed the regularity condition ${\tilde R}'_{\rm in}(0)=0$ at the center of the Sun and 
by convenience set to zero the integration constant at the exterior. In principle if we extrapolate the exterior solution to $r=\infty$ the  
solution $R(r)= {\tilde R}(r) + R_0$ reaches the cosmological value $R_0$ at spatial infinity. Here the mass $M_\odot= 4\pi \rho_\odot {\cal R}^{3}_\odot/3$. 
This solution agrees, for instance, with Eq.~(114) of Ref.~\cite{Sotiriou2010}. Notice that ${\tilde R}_{\rm out}(r)/R_0= \kappa M_\odot/(12\pi R_0 f_{RR_0} r)
= 2 G_0 M_\odot/(3 R_0 f_{RR_0} r) \sim G_0 M_\odot/r \sim \psi_{\rm out} \sim \phi_{\rm out}$. It is then expected that this solution will disturb considerably the 
solar-system tests, as we will show next. 

Using the interior solution ${\tilde R}_{\rm in}(r)$ in Eq.~(\ref{psiprimesun2}) taking $R_{\rm min}= R_0$
together with the condition (\ref{R0vac2}) we obtain
\begin{eqnarray}
&& \frac{1}{r}\frac{d}{dr}\left(r\psi_{\rm in}\right) = \frac{\kappa{\rho_\odot} r}{3 f_{R_0}}\left\{ 1 - \frac{1}{24}\Big{[}R_0 r^2 
\right.  \nonumber \\
&& - \left.\left. R_0 {{\cal R}_\odot}^2 \left(1 + \frac{2f_{R_0}}{R_0 f_{RR_0}}\right)\left(1-\frac{r^2}{3{{\cal R}_\odot}^2}\right)\right]\right\} \;.
\end{eqnarray}
By the arguments given previously, all the terms with $R_0 r^2= \Lambda_{\rm eff} r^2/4\ll 1$ and $R_0 {{\cal R}_\odot}^2= \Lambda_{\rm eff} {{\cal R}_\odot}^2/4 \ll 1$ 
in the neighborhood of the Sun. Moreover $\frac{f_{R_0}}{R_0 f_{RR_0}}$ are of order unity. Therefore with a very good approximation we have
\begin{equation}
\label{psiinfRw}
 \frac{1}{r}\frac{d}{dr}\left(r\psi_{\rm in}\right) \approx \frac{\kappa{\rho_\odot} r}{3 f_{R_0}}\;. \hspace{1.5cm}  (0\leq r \leq {\cal R}_\odot) 
\end{equation}
This equation coincides with Guo's Eq.~(19) when matching both notations, his and ours (cf. footnote~\ref{foot:notat}).
Comparing the r.h.s of this equation with the r.h.s of Eq.~(\ref{psiGR}) when taking $\rho=\rho_\odot$ for the interior solution, 
we appreciate that both differ by a factor $2/(3f_{R_0})$, which will be also manifested in the solution itself.

Equation (\ref{psiinfRw}) can be easily solved
\begin{equation}
\psi_{\rm in}(r)= \frac{\kappa M_\odot r^2}{12\pi  {{\cal R}_\odot}^3 f_{R_0}}= \frac{2 G_0 M_\odot r^2}{3  {{\cal R}_\odot}^3 f_{R_0}} \hspace{0.5cm} (0\leq r \leq {\cal R}_\odot)  \;,
\end{equation}
In a similar way, we obtain the exterior solution by using the exterior solution ${\tilde R}_{\rm out}(r)$ in Eq.~(\ref{psiprimesun2})
\begin{equation}
 \frac{1}{r}\frac{d}{dr}\left(r\psi_{\rm out}\right) = -\frac{\kappa{M_\odot} R_0}{288\pi f_{R_0}}\left( 1 - \frac{4f_{R_0}}{R_0f_{RR_0}}\right)  \;.
\end{equation}
Integrating we obtain
\begin{equation}
 \psi_{\rm out}(r) = \frac{const}{r} -\frac{\kappa{M_\odot}}{r} \frac{R_0 r^2}{576\pi f_{R_0}}\left( 1 - \frac{4f_{R_0}}{R_0f_{RR_0}}\right)  \;.
\end{equation}
where $const$ is an integration constant. We take this solution to be valid for $r\lesssim 150 \, {\rm A.U.}$ In this region the second term is 
very small compared with the first one since the leading term will be $\sim G_0M_\odot/r$ when matching with the interior solution. Finally we obtain
\begin{equation}
 \psi_{\rm out}(r) \approx \frac{\kappa M_\odot}{12\pi f_{R_0}}\frac{1}{r}= \frac{2 G_0 M_\odot}{3 f_{R_0}}\frac{1}{r} \hspace{1cm}  ({\cal R}_\odot \leq r) \;, 
\end{equation}
Comparing with ${\tilde R}_{\rm out}(r)$ from Eq.(\ref{solRtildew}), we notice $\psi_{\rm out}(r) = {\tilde R}_{\rm out}(r) f_{RR_0}/f_{R_0}= ({\tilde R}_{\rm out}(r)/R_0) (R_0f_{RR_0}/f_{R_0})\sim {\tilde R}_{\rm out}(r)/R_0$, 
as expected. In summary
\begin{equation}
\psi (r)\approx
\begin{cases}
 \frac{\kappa M_\odot r^2}{12\pi  {{\cal R}_\odot}^3 f_{R_0}}= \frac{2 G_0 M_\odot r^2}{3  {{\cal R}_\odot}^3 f_{R_0}} \hspace{1cm} (0\leq r \leq {\cal R}_\odot)\\
  \\
 \frac{\kappa M_\odot}{12\pi f_{R_0}}\frac{1}{r}= \frac{2 G_0 M_\odot}{3 f_{R_0}}\frac{1}{r} \hspace{.3cm}  ({\cal R}_\odot \leq r \lesssim 150 {\rm A.U.}) \;.
\end{cases}
\label{solpsiw}
\end{equation}

We stress again that this solution is continuous at ${\cal R}_\odot$ but its derivative is not defined there. The exterior solution coincides, for instance, 
with Eq.~(131) of Ref.~\cite{Sotiriou2010}.

As concerns the solution for $\phi$, instead of using Eq.~(\ref{phiprime2}) we use Eq.~(\ref{psimphi}). By the same arguments given before, all the terms that involve 
$R_0= \Lambda_{\rm eff}$ will provide, once the solution of ${\tilde R}(r)$ is replaced, terms containing dimensionless factors $R_0 r^2$ which are very small 
compared with the rest of the terms. So we shall neglect them, as well as the pressure, and obtain
\begin{eqnarray}
\label{psimphiaprox}
&& \frac{d}{dr}\left(\psi-\phi +  \frac{{\tilde R} f_{RR_0}}{f_{R_0}}\right) \approx \frac{\kappa r }{2f_{R_0}}\left(T^{r}_{\,\,r}- T^{t}_{\,\,t} + \frac{T}{3}\right)
\nonumber \\
&& \approx \frac{\kappa \rho r }{3f_{R_0}}
\end{eqnarray}
We remind the reader that in GR we have instead Eq.(\ref{psimphiGR}). The interior solution of Eq.~(\ref{psimphiaprox}) reads then
\begin{equation}
\phi_{\rm in}(r)= \psi_{\rm in}(r) + \frac{{\tilde R}_{\rm in}(r) f_{RR_0}}{f_{R_0}}- \frac{\kappa \rho r^2 }{6f_{R_0}} + {\rm const} \;,
\end{equation}
which is valid for $0\leq r \leq {\cal R}_\odot$ and where we introduced an integration constant. When replacing the interior solutions given by 
Eq.~(\ref{solpsiw}) and Eq.~(\ref{solRtildew}) we obtain
\begin{equation}
\phi_{\rm in}(r)= {\rm const} - \frac{\kappa M_\odot r^2}{24\pi  {{\cal R}_\odot}^3 f_{R_0}} + \frac{\kappa M_\odot}{8\pi  {\cal R}_\odot f_{R_0}}\left(1-\frac{r^2}{3 {{\cal R}_\odot}^2} \right)
\;.
\end{equation}
The integration constant will be determined when matching the interior and the exterior solutions at $r={\cal R}_\odot$.

The exterior solution in vacuum is given by:
\begin{equation}
\phi_{\rm out}(r)\approx \psi_{\rm out}(r) + \frac{{\tilde R}_{\rm out}(r) f_{RR_0}}{f_{R_0}} + {\rm const} \;.
\end{equation}
For $r\sim 150 \, {\rm A.U.}$, $\phi_{\rm out} \sim 0$, so we can neglect the integration constant. 
When replacing the exterior solutions given by Eq.~(\ref{solpsiw}) and Eq.~(\ref{solRtildew}) yields
\begin{equation}
\phi_{\rm out}(r)\approx  \frac{\kappa M_\odot}{6\pi f_{R_0}}\frac{1}{r}= \frac{4 G_0 M_\odot}{3 f_{R_0}}\frac{1}{r} \hspace{1cm}  ({\cal R}_\odot \leq r) \;.
\end{equation}
Matching $\phi_{\rm in}({\cal R}_\odot)= \phi_{\rm out}({\cal R}_\odot)$ we obtain the complete solution which is summarized as follows,
\begingroup\makeatletter\def\f@size{9}\check@mathfonts
\def\maketag@@@#1{\hbox{\m@th\large\normalfont#1}}%
\begin{equation}
\phi (r)\approx
\begin{cases}
  \frac{\kappa M_\odot}{4\pi  {{\cal R}_\odot} f_{R_0}}\left(1 - \frac{r^2}{3{{\cal R}_\odot}^2}\right)= \frac{2G_0 M_\odot}{{{\cal R}_\odot} f_{R_0}}\left(1 - \frac{r^2}{3{{\cal R}_\odot}^2}\right) \\
  \hspace{4.3cm} (0\leq r \leq {\cal R}_\odot) \\
 \\
  \frac{\kappa M_\odot}{6\pi f_{R_0}}\frac{1}{r}= \frac{4 G_0 M_\odot}{3 f_{R_0}}\frac{1}{r} \hspace{1.3cm}  ({\cal R}_\odot \leq r \lesssim 150 \, {\rm A.U.} ) \;.
\end{cases}
\label{solphiw}
\end{equation}
\endgroup

The metric perturbation $\phi(r)$ and its derivative are both continuous at $r={\cal R}_\odot$, unlike $\psi(r)$, which is continuous
but not its derivative.

The solutions Eq.~(\ref{solpsiw}) and~(\ref{solphiw}) coincide with the exterior solutions of Guo's Ref.~\cite{Guo2014}\footnote{\label{foot:notat}In Ref.~\cite{Guo2014} 
a different notation for the metric is used, so to match the notations one should use $1-2\phi= N(r)$ and $1+2\psi= 1/B(r)$. For the latter, 
$B(r)\approx 1-2\psi$.}. The interior solution $\psi_{\rm in}(r)$ also agrees with Guo's $B(r)$, 
however, there is a discrepancy between our $\phi_{\rm in}(r)$ and Guo's  interior solution for $N(r)$. Furthermore, the interior Schwarzschild solution of GR at order $r^2$
agrees with our interior GR solution $\phi_{\rm in}(r)$, but it does not agree with Guo's interior solution~\footnote{From Eq.(\ref{psimphiGR}) we obtain
\begin{eqnarray}
\phi_{\rm in}(r) &=& \psi_{\rm in}(r) - \frac{\kappa \rho_\odot}{4}\left( r^2- {{\cal R}_\odot}^2\right) \nonumber \\
&=& \psi_{\rm in}(r) + \frac{3 \kappa M_\odot}{16\pi {{\cal R}_\odot}}\left(1-\frac{r^2}{{{\cal R}_\odot}^2}\right) \nonumber \\
&=& \frac{G_0  M_\odot r^2}{{{\cal R}_\odot}^3} + \frac{3 G_0M_\odot}{2 {{\cal R}_\odot}}\left(1-\frac{r^2}{{{\cal R}_\odot}^2}\right) \nonumber \\
&=& \frac{3 G_0M_\odot}{2 {{\cal R}_\odot}}\left(1-\frac{r^2}{3{{\cal R}_\odot}^2}\right)
\end{eqnarray}
Thus, $-g_{tt}= n(r)= 1-2\phi\approx 1- \frac{3 G_0M_\odot}{{{\cal R}_\odot}}\left(1-\frac{r^2}{3{{\cal R}_\odot}^2}\right)$ which agrees with the 
the interior Schwarzschild solution in GR at order $r^2$. Guo's interior solution reads
$-g_{tt}= N(r)= 1- \frac{6\epsilon}{r_0} + \frac{3\epsilon}{r_0} (\frac{r}{r_0})^2= 1 - \frac{6\epsilon}{r_0}\left(1-\frac{r^2}{2{r_0}^2}\right) = 1 - \frac{4G_0 M}{r_0}\left(1-\frac{r^2}{2{r_0}^2}\right)$. In the case of the Sun, $r_0$ stands for ${\cal R}_\odot$ and $M$ for $M_\odot$.  So even if this solution correctly matches the exterior solution at $r=r_0$, it does not agree with our solution in the interior neither does it recover the solution of GR.}

Nevertheless, these differences do not change the conclusions, since the relevant quantities to compare with observations are the exterior ones. In any case, and as the title of this Appendix indicates, this analysis is flawed as it turns to be inconsistent with
observations. We see that the PNP $\gamma= \psi_{\rm out}/\phi_{\rm out}\approx 1/2$ independently of the non-linear model $f(R)$. 
Since the observations indicate $|\gamma-1|\lesssim 2.3 \times 10^{-5}$, thus, this naive analysis implies that
all non-linear $f(R)$ models are blatantly ruled-out by 
about four orders of magnitude. Nonetheless, the main goal of the paper was to show that this (naive) conclusion is actually wrong. In the above linear analysis we completely neglected the screening effects by perturbing the scalar DOF only around the cosmological background, instead around the
minima inside and outside the Sun. This amounts to neglect the equivalent potential resulting when treating $f(R)$ theory as a kind of 
Brans-Dicke (or chameleon) theory without the kinetic term for the field $\chi$, i.e., with $\omega_{\rm BD}\equiv 0$. The naive Brans-Dicke approach leads then to
$\gamma=1/2$, as well.

\section{Robustness of the quadratic approximation in the effective potential}
\label{appA}
 
In Section~\ref{sec:QA}, Eq.~(\ref{traceRrap}) was solved using a linearization of the effective potential around its minimum in each region of the solar system. However, this approximation is not always the most appropriate. For example, in dense regions it may happen that $\frac{\kappa\rho}{3f_{RR}}\gg\frac{2 f - R f_{R}}{3 f_{RR}}$, in which case a more suited approximation to Eq.~(\ref{traceRrap}) is

\begin{equation}
\label{TSR}  
\nabla^2R=\frac{-\kappa\rho}{3f_{RR}}\;.
\end{equation}

So, in order to determine which  of the two approximations is better, we proceed with the following analysis  which is inspired by the development presented 
in \cite{W06}. 
\begin{eqnarray}
  \frac{dV_{\rm lin}}{dR} &=& m_{\rm eff}^2(R-R_{\rm min}) \;,\\
\frac{dV_{\rm mat}}{dR} &=& \frac{-\kappa\rho}{3f_{RR}}  \;.
\end{eqnarray}

\begin{figure}[t]
\centering
\includegraphics[width=0.52\textwidth]{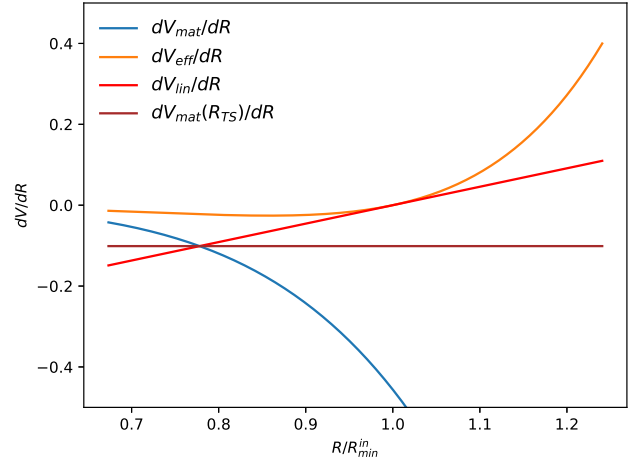}
\caption{Behavior of the terms involved in the potential's derivative and the different linear approximations considered (in units of 
$10^{91}$ {\rm cm}$^{-4}$) as a function of $R/R_{\rm min}^{\rm in}$ within the Sun. Here the plots correspond to the Starobinsky model (\ref{Staromod}) with $q=2$ 
but a similar behavior can be seen for other $f(R)$ models.}
\label{fig:dVdR_adentro}  
\end{figure}

We call $R_{TS}$ the value of $R$  for which
\begin{equation}
\frac{dV_{\rm lin}}{dR} =  \frac{dV_{\rm mat}}{dR} \;.
\label{RTS}
\end{equation}
Figure~\ref{fig:dVdR_adentro} shows that if  $R<R_{TS}$,  $\frac{dV_{\rm mat}}{dR}$ is a good approximation to the derivative of the effective potential and therefore Eq.~(\ref{traceRrap}) can be approximated by Eq.~(\ref{TSR})  while if $R>R_{TS}$ the linear approximation to the derivative of the effective potential Eq.~(\ref{eqRL}) is the appropriate one.

Let us  consider at this point a symmetric spherical body with density $\rho$ and radius ${\cal{R}}_c$, surrounded by a region of lower density $\rho_{\rm out}$.   If $R(r)$ is the solution of Eq.(\ref{TSR})  throughout the region of density $\rho$, it must satisfy
\begin{equation}
m_{\rm eff}^2(R(0)-R_{\rm min})<\frac{-\kappa\rho}{3f_{RR}}\;,
\label{meff4}
\end{equation}  
and it is useful to remember that $R(r)$ reaches its maximum value at the center of the body. 
Next we develop the previous expression to obtain a condition that depends only on the minima of the 
effective potential and the physical parameters of the problem.
Unfortunately, for most of the $f(R)$ models Eq.(\ref{TSR}) cannot be solved analytically nor its r.h.s can be linearized. However, according to Figure \ref{fig:dVdR_adentro}, $\frac{dV_{\rm mat}}{dR}$ can be well approximated by $\frac{dV_{\rm mat}(R_{TS})}{dR}$ when $R<R_{TS}$, which is the region where Eq. (\ref{meff4}) holds.  Therefore, we replace $f_{RR}$ with $f_{RR}|_{R_{TS}}=f_{RR_{TS}}$ in Eq.~(\ref{TSR}). 

Considering now the case of the Sun surrounded by the interstellar medium, the inequality (\ref{meff4}) can be rewritten as:
\begin{eqnarray}
&& R_{\rm min}^{\rm IM}-R_{\rm min}^{\rm in}+\frac{\kappa\rho_{\odot}\mathcal{R}_{\odot}^2}{18f_{RR_{\rm TS}}}\left(1+\frac{2}{1+m^{\rm IM}_{\rm eff}\mathcal{R}_{\odot}}\right)
\nonumber \\
&& <\frac{-\kappa\rho_{\odot}}{3\left(m^{\rm in}_{\rm eff}\right)^2f_{RR_{\rm TS}}},
\label{condTS}
\end{eqnarray}

Consequently, when the above condition is not satisfied, the solution  that results from linearizing the effective potential  around its minimum (the one used in this paper to solve Eq.(\ref{traceRrap})) is a better approximation. Moreover, when (\ref{condTS}) is {\it not} satisfied, but its opposite is,
  we have instead the following condition written in dimensionless form: 
\begin{equation}
  \frac{3(R_{\rm min}^{\rm in}-R_{\rm min}^{\rm IM})f_{RR_{\rm TS}}}{\Phi_{\odot}}\leq\frac{6}{\left(m^{\rm in}_{\rm eff}\mathcal{R}_{\odot}\right)^2}+1+\frac{2}{1+m^{\rm IM}_{\rm eff}\mathcal{R}_{\odot}},
\label{condTS2}
\end{equation}  
where $\Phi_{\odot}=\kappa\rho_{\odot}\mathcal{R}_{\odot}^2/6=G_0M_{\odot}/\mathcal{R}_{\odot}$ is the Newtonian potential associated with the Sun. 
Furthermore,  if $m^{\rm IM}_{\rm eff}\mathcal{R}_{\odot}\ll1$ and $m^{\rm in}_{\rm eff}\mathcal{R}_{\odot}\gg 1$, the condition (\ref{condTS2}) becomes
\begin{equation}
  \frac{(R_{\rm min}^{\rm in}-R_{\rm min}^{\rm IM})f_{RR_{\rm TS}}}{\Phi_{\odot}}\lesssim 1\;.
  \label{ineq1}
\end{equation}
Meanwhile, if $m^{\rm IM}_{\rm eff}\mathcal{R}_{\odot}\sim 1$, the r.h.s. of (\ref{ineq1}) is replaced by $2/3$.
Finally, if $m^{\rm IM}_{\rm eff}\mathcal{R}_{\odot}\gg 1$, the r.h.s. changes to $1/3$.

On the other hand, if like in Sec.~\ref{sec:QA} we consider a three-region configuration 
(Sun, the corona and IM) and after some considerations such as $m^{\rm cor}_{\rm eff}\gg m^{\rm IM}_{\rm eff}$ and $R_{\rm min}^{\rm cor}\gg R_{\rm min}^{\rm IM}$, the condition (\ref{condTS}) becomes 
\begin{eqnarray}
 R_{\rm min}^{\rm cor}-R_{\rm min}^{\rm in}&+&\frac{\kappa\rho_{\odot}\mathcal{R}_{\odot}^2}{18f_{RR_{\rm TS}}}\left(1+\frac{2}{1+m^{\rm cor}_{\rm eff}\mathcal{R}_{\odot}}\right) \nonumber\\
 &&  <\frac{-\kappa\rho_{\odot}}{3\left(m^{\rm in}_{\rm eff}\right)^2f_{RR_{\rm TS}}}\;.
 \label{ineq2}
\end{eqnarray}
Therefore, the solution that results from linearizing the effective potential around its minimum is a better approximation when the opposite
  of inequality (\ref{ineq2}) holds, that is, when the following condition is satisfied
\begin{equation}
  \frac{18f_{RR_{\rm TS}}}{\kappa\rho_{\odot}\mathcal{R}_{\odot}^2}(R_{\rm min}^{\rm in}-R_{\rm min}^{\rm cor})
  \leq\frac{6}{\left(m^{\rm in}_{\rm eff}\mathcal{R}_{\odot}\right)^2}+1+\frac{2}{1+m^{\rm cor}_{\rm eff}\mathcal{R}_{\odot}}\;.
\end{equation}
When $m_{\rm eff}^{\rm in}{\cal{R}_\odot} \gg 1 \gg m_{\rm eff}^{\rm cor} {\cal{R}_\odot}$, the above condition can be approximated by
\begin{equation}
  \frac{(R_{\rm min}^{\rm in}-R_{\rm min}^{\rm cor})f_{RR_{\rm TS}}}{G_0M_{\odot}/\mathcal{R}_{\odot}}\lesssim 1  \;,
\label{condTS4}
\end{equation}
but if $m_{\rm eff}^{\rm cor}  {\cal{R}_\odot} \sim  1$, the r.h.s of (\ref{condTS4}) is replaced by 2/3; whereas
if $m_{\rm eff}^{\rm cor} {\cal{R}_\odot} \gg 1$, the r.h.s changes to 1/3.


\end{document}